\documentclass[opre,nonblindrev]{informs3}
\usepackage{amsmath}
\usepackage{cleveref}
\usepackage{algpseudocode}
\usepackage{algorithm}
\usepackage{caption}
\usepackage{subcaption}

\DoubleSpacedXI 

\usepackage{endnotes}
\let\footnote=\endnote

\usepackage{natbib}
 \bibpunct[, ]{(}{)}{,}{a}{}{,}%
 \def\bibfont{\small}%
\TheoremsNumberedThrough     
\ECRepeatTheorems

\EquationsNumberedThrough    

\begin{document}


\RUNAUTHOR{Lam and Mottet}

\RUNTITLE{Worst-case Tail Analysis}

\TITLE{Tail Analysis without Parametric Models: A Worst-case Perspective}

\ARTICLEAUTHORS{%
\AUTHOR{Henry Lam}
\AFF{Department of Industrial Engineering and Operations Research, Columbia University, New York, NY 10027, \EMAIL{henry.lam@columbia.edu}} 
\AUTHOR{Clementine Mottet}
\AFF{Department of Mathematics and Statistics, Boston University, Boston, MA 02215, \EMAIL{cmottet@bu.edu}}
} 

\ABSTRACT{%
A common bottleneck in evaluating extremal performance measures is that, due to their very nature, tail data are often very limited. The conventional approach selects the best probability distribution from tail data using parametric fitting, but the validity of the parametric choice can be difficult to verify. This paper describes an alternative based on the computation of worst-case bounds under the geometric premise of tail convexity, a feature shared by all common parametric tail distributions. We characterize the optimality structure of the resulting optimization problem, and demonstrate that the worst-case convex tail behavior is in a sense either extremely light-tailed or extremely heavy-tailed. We develop low-dimensional nonlinear programs that distinguish between the two cases and compute the worst-case bound. We numerically illustrate how the proposed approach can give more reliable performances than conventional parametric methods.


}%



\maketitle

%


\title{Tail Analysis without Parametric Models: A Worst-case Perspective}
\author{}
\date{}
\maketitle

\section{Introduction}\label{sec:intro}
Modeling extreme behaviors is a fundamental task in analyzing and managing risk. As the earliest applications, hydrologists and climatologists study historical data of sea levels and air pollutants to estimate the risk of flooding and pollution (\cite{gumbel2012statistics}). In non-life or casualty insurance, insurers rely on accurate prediction of large losses to price and manage insurance policies (\cite{McNeil1997,beirlant1992modeling,embrechts1997modelling}). Relatedly, financial managers estimate risk measures of portfolios to safeguard losses (\cite{glasserman2005importance,glasserman2007large,glasserman2008fast}). In engineering, measurement of system reliability often involves modeling the tail behaviors of individual components' failure times (\cite{nicola1993fast,heidelberger1995fast}). 

Despite its importance in various disciplines, tail modeling is an intrinsically difficult task because, by their own nature, tail data are often very limited. 
Consider these two examples:
\begin{example}[Adopted from \cite{McNeil1997}]
\emph{There were $2,156$ Danish fire losses over one million Danish Krone (DKK) from $1980$ to $1990$. The empirical cumulative distribution function (ECDF) and the histogram (in log scale) are plotted in Figure \ref{Danish fire loss}. For a concrete use of the data, an insurance company might be interested in pricing a high-excess contract with reinsurance, which has a payoff of $X-50$ (in million DKK) when $50<X\leq200$, $150$ when $X>200$, and $0$ when $X\leq50$, where $X$ is the loss amount (the marks $50$ and $200$ are labeled with vertical lines in Figure \ref{Danish fire loss}). Pricing this contract would require, among other information, $E[\text{payoff}]$. However, only seven data points are above $50$ (the loss amount above which the payoff is non-zero).}\label{example:fire}
\end{example}


\begin{centering}
\begin{minipage}{\textwidth}
  \begin{minipage}[b]{0.47\textwidth}
    \centering
    \includegraphics[scale=.22]{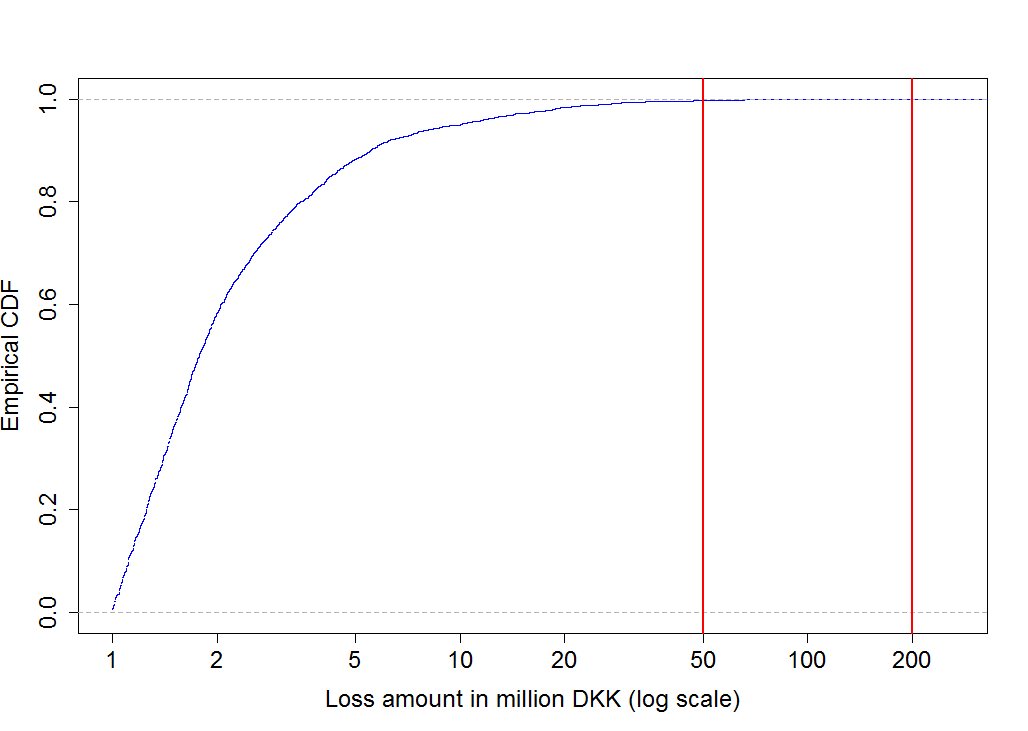}
  \end{minipage}
  \hspace{.3cm}
  \begin{minipage}[b]{0.47\textwidth}
    \centering
    \includegraphics[scale=.22]{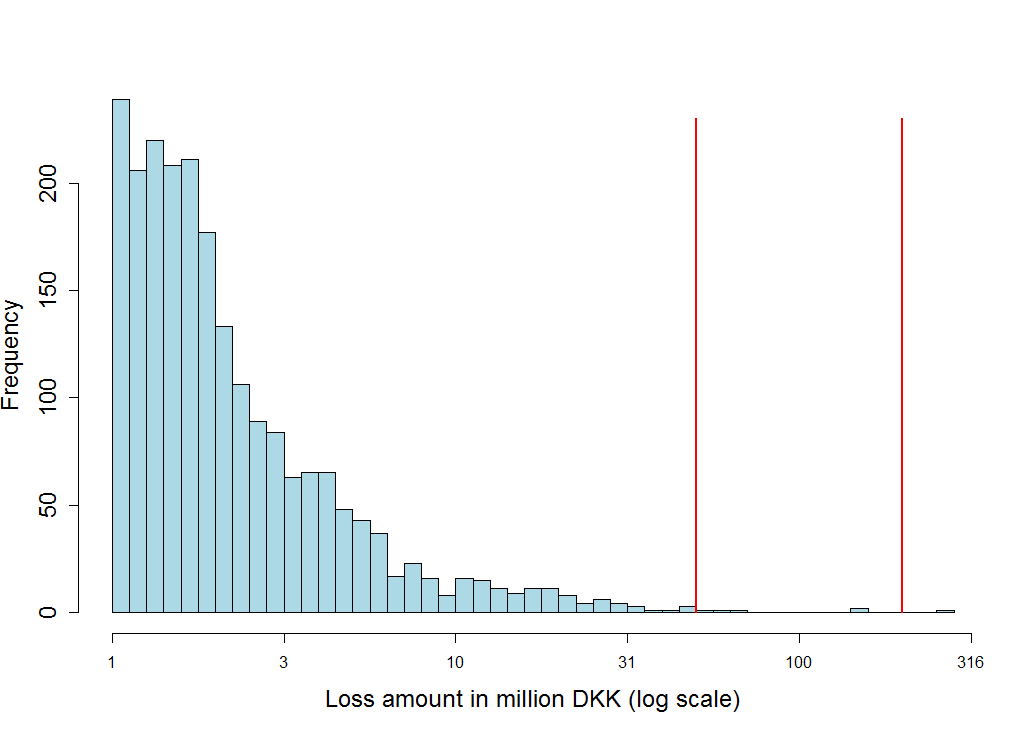}
  \end{minipage}
  \captionof{figure}{ECDF and histogram for Danish fire losses from $1980$ to $1990$}
  \end{minipage}\label{Danish fire loss}
  \end{centering}

\begin{figure}[h]
\centering
\includegraphics[scale = 0.25]{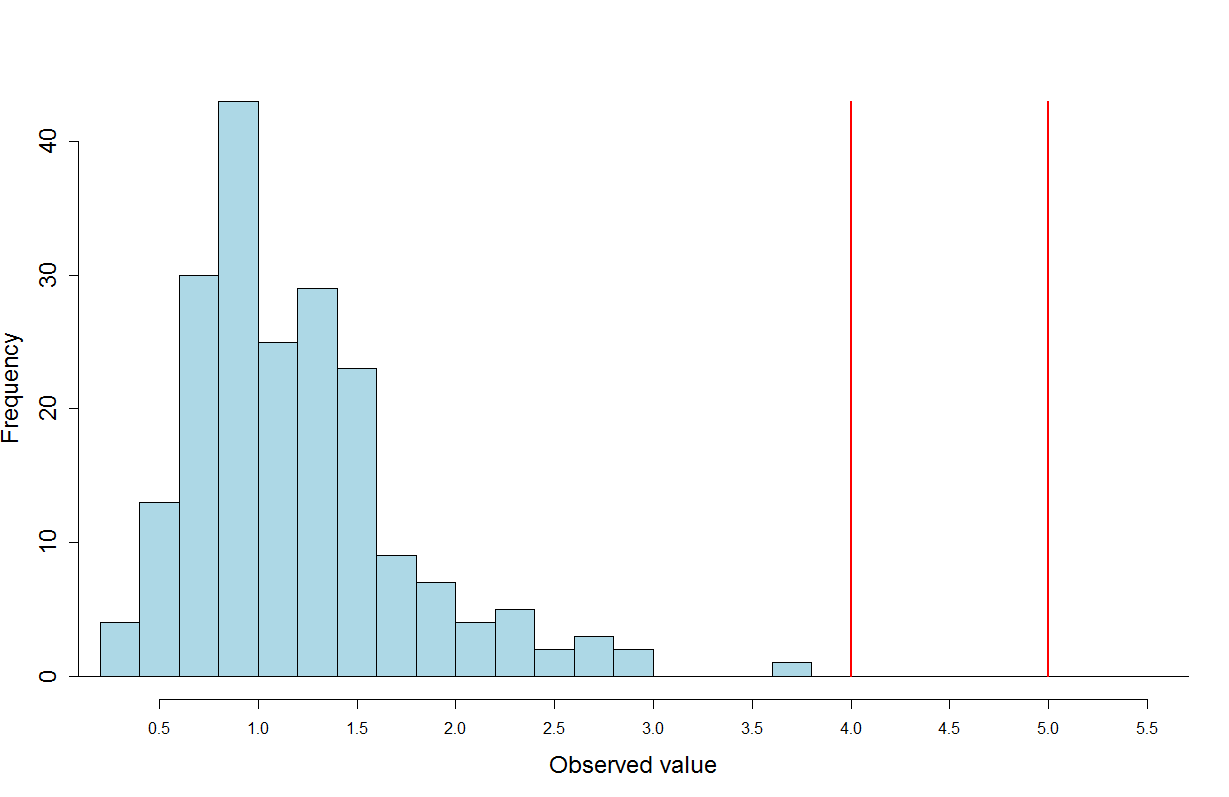}
\caption{Histogram of a synthetic data set with sample size $200$}
\label{synthetic data}
\end{figure}

\begin{example}
\emph{A more extreme situation is a synthetic data set of size $200$ generated from an unknown distribution, whose histogram is shown in Figure \ref{synthetic data}. Suppose the quantity of interest is $P(4<X<5)$. This appears to be an ill-posed problem since the interval $[4,5]$ has no data at all. This situation is not uncommon when in any application one tries to extrapolate the tail with a small sample size.}\label{example:synthetic}
\end{example}

The purpose of this paper is to develop a theoretically justified methodology to estimate tail-related quantities of interest such as those depicted in the examples above. This requires drawing information properly from data not in the tail. We will illustrate how to do this and revisit the two examples later with numerical performance of our method.


\section{Our Approach and Main Contributions}
We adopt a nonparametric approach. 
Rather than fitting a tail parametric curve when there can be few or zero observations in the tail region, we base our analysis on the geometric premise that the tail density is convex. We emphasize that this condition is satisfied by \emph{all} common parametric distributions (e.g. normal, lognormal, exponential, gamma, Weibull, Pareto etc.). 
For this reason we believe it is a natural and minimal assumption to make.  

In any given problem, there can be potentially infinitely many feasible candidates of convex tails. The central idea of our method is a worst-case characterization. Formally, given information on the non-tail part of the distribution and a target quantity of interest (e.g., $P(4<X<5)$ in Example \ref{example:synthetic}), we aim to find a convex tail, consistent with the non-tail part, that gives rise to the worst-case value of the target (e.g., the largest possible value of $P(4<X<5)$). This value serves as a tight bound for the target that is robust with respect to the ambiguity of the tail, without using any particular tail knowledge other than our a priori assumption of convexity.

Our proposed approach requires solving an optimization over a potentially infinite-dimensional space of convex tails. As our key contributions, we show that this problem has a very simple optimality structure, and find its solution via low-dimensional nonlinear programs. 
In particular:
\begin{enumerate}
\item We characterize the worst-case tail behavior under the tail convexity condition. We show that the worst-case tail, for \emph{any} bounded target quantity of interest, is in a sense either \emph{extremely light-tailed or extremely heavy-tailed}. Both cases can be characterized by piecewise linear densities, the distinction being whether the pieces form a bounded support distribution or lead to probability masses that escape to infinity.

\item We provide efficient algorithms to distinguish between the two cases above, and to solve for the optimal distribution in each case. For a large class of objectives, the algorithm requires at most a two-dimensional nonlinear program. 
\end{enumerate}

Our approach outputs statistically valid worst-case bounds when integrating with confidence estimates drawn from the non-tail portion of the data. This approach uses the convexity assumption to get around the difficulty faced by conventional parametric methods (discussed in detail in the next section) in directly estimating the tail curve, by effectively mitigating the estimation burden to the central part of the density curve where more data are available. However, we pay the price of conservativeness: our method can generate a worst-case bound that is over-pessimistic. We therefore believe it is most suitable for small sample size, when a price of conservativeness is unavoidable in trading with statistical validity.

The remainder of this paper is organized as follows. Section \ref{sec:literature techniques} discusses some previous techniques and reviews the relevant literature. Section \ref{sec:basic} presents our formulation and results for an abstract setting. Section \ref{sec:opt procedure} studies the numerical solution algorithm. Section \ref{sec:numerics data} focuses on integrating these results with data. Section \ref{sec:numerics} shows some numerical illustration. Section \ref{sec:discussions} concludes and discusses future work. Some auxiliary theorems and proofs are left to the Appendix.

\section{Related Work}\label{sec:literature techniques}
\subsection{Overview of Common Tail-fitting Techniques}\label{Sec:Existing Techniques}
As far as we know, all existing techniques for modeling extreme events are parametric-based, in the sense that a ``best" parametric curve is chosen and the parameters are fit to the tail data. The classic text of \cite{hogg2009loss} provides a comprehensive discussion on the common choices of parametric tail densities. While exploratory data analysis, such as quantile plots and mean excess plots, can provide guidance regarding the class of parametric curves to use (such as heavy, middle or light tail), this approach is limited by its reliance on a large amount of data in the tail and subjectivity in the choice of parametric curve.

Beyond the goodness-of-fit approach, there are two widely used results on the parametric choice that is provably suitable for extreme values. The Fisher-Tippett-Gnedenko Theorem (\cite{fisher1928limiting,gnedenko1943distribution}) postulates that the sample maxima, after suitable scaling, must converge to a generalized extreme value (GEV) distribution, given that it converges at all to some non-degenerate distribution. 
This result is useful if the data are known to derive from the maximum of some distributions. For instance, environmental data on sea level and river heights are often collected as annual maxima (\cite{davison1990models}), and in this scenario it is sensible to fit the GEV distribution. In other scenarios, the data have to be pre-divided into blocks and blockwise maxima have to be taken in order to apply GEV, but this blockwise approach is statistically wasteful (\cite{embrechts2005quantitative}).

The Pickands-Balkema-de Haan Theorem (\cite{pickands1975statistical,balkema1974residual}) does not require data to come from maxima. Rather, the theorem states that the excess losses over thresholds converge to a generalized Pareto distribution (GPD) as the thresholds approach infinity, under the same conditions as the Fisher-Tippett-Gnedenko Theorem. 
The Pickands-Balkema-de Haan theorem provides a solid mathematical justification for using GPD to fit the tail portion of data (\cite{McNeil1997,embrechts2005quantitative}). Fitting GPD can be done by well-studied procedures such as maximum likelihood estimation (\cite{smith1985maximum}), and the method of probability-weighted moments (\cite{hosking1987parameter}). The Hill estimator (\cite{hill1975simple,davis1984tail}) is also a widely used alternative.

Despite the attraction and frequent usage, fitting GPD suffers from two pitfalls: First, there is no convergence rate result that tells how high a threshold should be for the GPD approximation to be valid (e.g. \cite{McNeil1997}). Hence, picking the threshold is an ad hoc task in practice. Second, and more importantly, even if the threshold chosen is sufficiently high for the approximation to hold, a large amount of data above it is needed to accurately estimate the parameters in GPD. In our two examples, especially Example \ref{example:synthetic}, this is plainly impossible.


\subsection{Related Literature on our Methodology}\label{sec:literature}
Our mathematical formulation and techniques are related to two lines of literature. The use of convexity and other shape constraints (such as log-concavity) have appeared in density estimation (\cite{cule2010maximum,seregin2010nonparametric,koenker2010quasi}) and convex regression (\cite{seijo2011nonparametric,hannah2013multivariate,lim2012consistency}) in statistics. A major reason for using convexity in these statistical problems is the removal of tuning parameters, such as bandwidth, as required by other methods such as the use of kernel. 

The second line of related literature is optimization over probability distributions, which have appeared in decision analysis (\cite{smith1995generalized,bertsimas2005optimal,popescu2005semidefinite}), robust control theory (\cite{iyengar2005robust,el2005robust,petersen2000minimax,hansen2008robustness}), distributionally robust optimization (\cite{delage2010distributionally,goh2010distributionally}), and stochastic programming (\cite{birge1987computing,birge1991bounding}). The typical formulation involves optimization of some objective governed by a probability distribution that is partially specified via constraints like moments (\cite{karr1983extreme,winkler1988extreme}) and statistical distances (\cite{ben2013robust}). Our formulation differs from these studies by its pertinence to tail modeling (i.e., knowledge of certain regions of the density, but none beyond it). Among all the previous works, only \cite{popescu2005semidefinite} has considered convex density assumption, as an instance of a proposed class of geometric conditions that are added to moment constraints. While the result bears similarity to ours in that a piecewise linearity structure shows up in the solution, our qualitative classification of the tail, the solution techniques, and the formulation in integrating with data all differ from the semidefinite programming approach in \cite{popescu2005semidefinite}. 

\section{Abstract Formulation and Results}\label{sec:basic}
We begin by considering an abstract formulation assuming full information on the distribution up to some threshold, and no information beyond. The next sub-sections give the details.


\subsection{Formulation}
Consider a continuous probability distribution on $\mathbb R$ whose density exists and is denoted by $f(x)$. We assume that $f$ is known up to a certain large threshold, say $a\in\mathbb R$. The goal is to extrapolate $f$.

We impose the assumption that $f(x)$, for $x\geq a$, is convex. Figure \ref{abstract} shows an example of an $f(x)$ known up to $a$, and Figures \ref{abstract convex} and \ref{abstract non-convex} each show an example of convex and non-convex extrapolation. Observe that the convex tail assumption excludes any ``surprising" bumps (and falls) in the density curve. 


\begin{centering}
\begin{minipage}{\textwidth}
  \begin{minipage}[b]{0.47\textwidth}
    \centering
    \includegraphics[scale=.24]{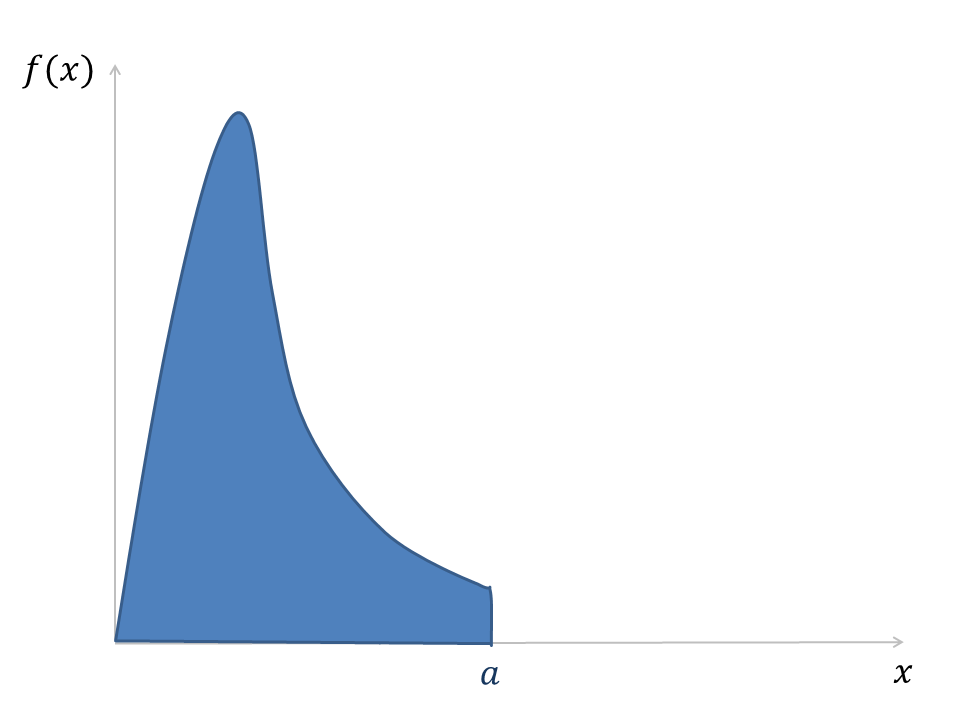}
    \captionof{figure}{A probability density $f(x)$ known up to a threshold $a$}
    \label{abstract}
  \end{minipage}
  \hspace{.3cm}
  \begin{minipage}[b]{0.47\textwidth}
    \centering
    \includegraphics[scale=.24]{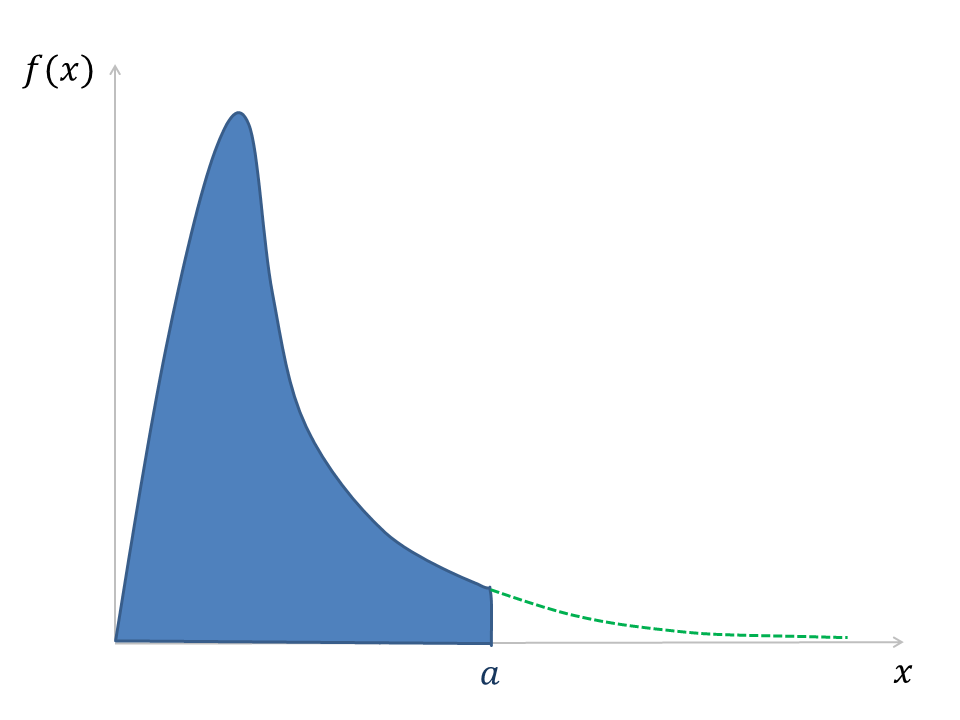}
      \captionof{figure}{An example of convex tail extrapolation}
      \label{abstract convex}
  \end{minipage}
    \hspace{.3cm}
  \begin{minipage}[b]{0.47\textwidth}
    \centering
    \includegraphics[scale=.24]{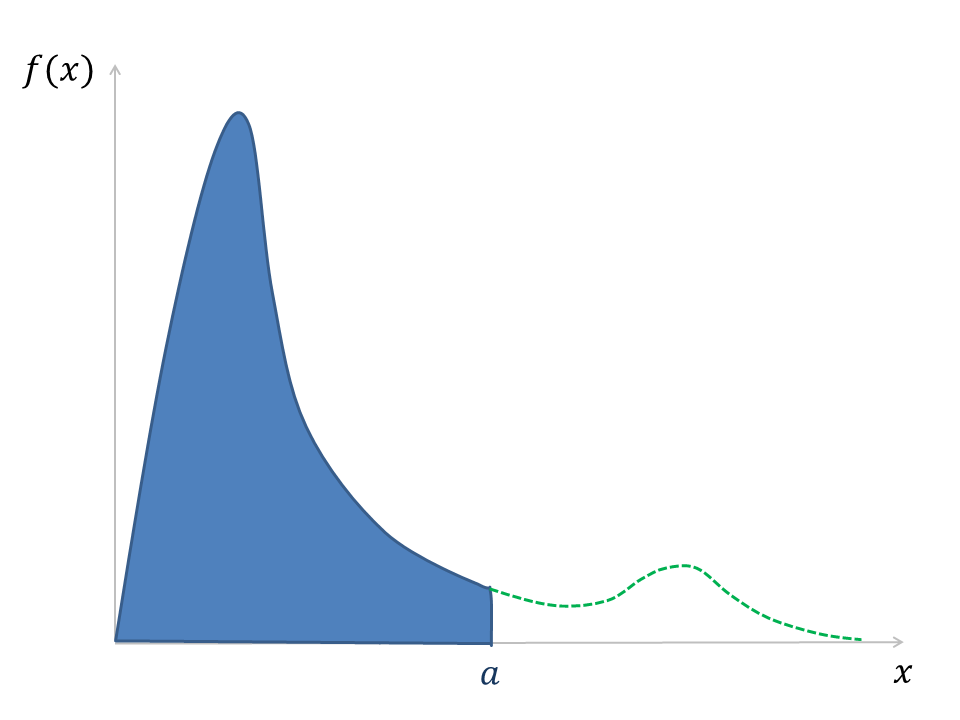}
     \captionof{figure}{An example of non-convex tail extrapolation}
     \label{abstract non-convex}
  \end{minipage}
    \hspace{.3cm}
    \begin{minipage}[b]{0.47\textwidth}
    \centering
    \includegraphics[scale=.27]{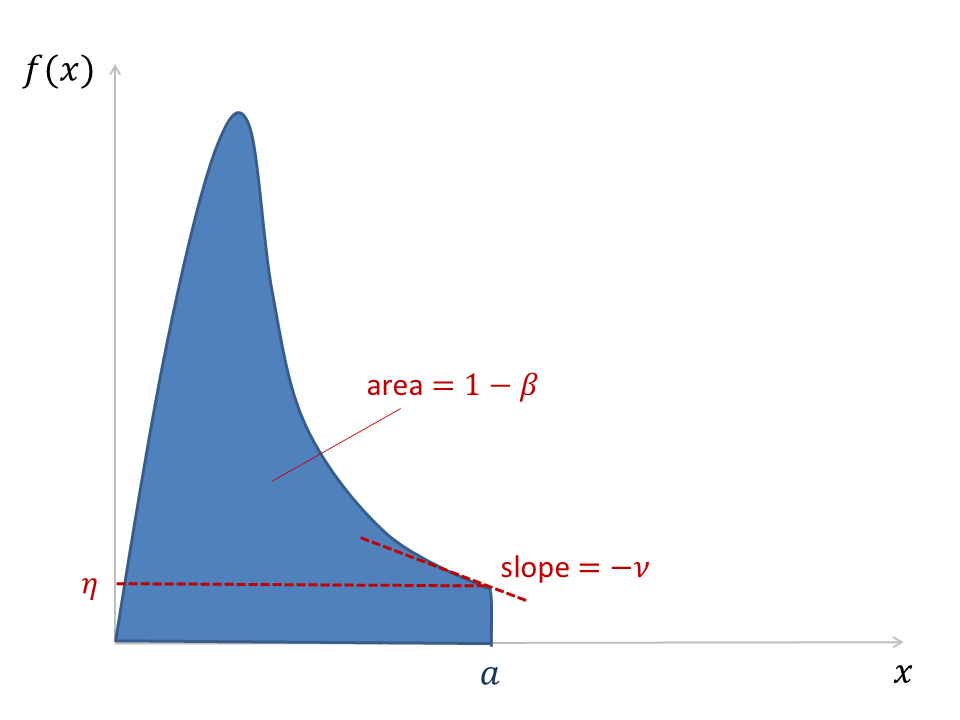}
     \captionof{figure}{The parameters $\eta,\nu,\beta$}
     \label{abstract parameters}
  \end{minipage}
    \end{minipage}
  \end{centering}

Now suppose we are given a target objective or performance measure $E[h(X)]$, where $E[\cdot]$ denotes the expectation under $f$, and $h:\mathbb R\to\mathbb R$ is a bounded function in $X$. The goal is to calculate the worst-case value of $E[h(X)]$ under the assumption that $f$ is convex beyond $a$. That is, we want to obtain $\max E[h(X)]=\int_{-\infty}^\infty h(x)f(x)dx$ where the maximization is over all convex $f(x),x\geq a$ such that it satisfies the properties of a probability density function. We assume that the density is left-differentiable at $a$, so that a convex extrapolation at $a$ can be suitably defined. For the formulation, we need three constants extracted from $f(x),x<a$, which we denote as $\eta,\nu,\beta>0$ respectively:
\begin{enumerate}
\item $\eta$ is the value of the density $f$ at $a$, i.e., $f(a)=\eta$.
\item $-\nu$ is the left derivative of $f$ at $a$, i.e., $f_-'(a)=-\nu$. We impose the condition that the right derivative $f_+'(a)\geq f_-'(a)=-\nu$. Note that, since $f$ is convex (and bounded) on $[a,\infty)$, its one-sided derivative exists everywhere on $[a,\infty)$ (\cite{rockafellar2015convex} Theorem 23.1).
\item $\beta$ is the tail probability at $a$. Since $f$ is known up to $a$, $\int_{-\infty}^a f(x)$ is known to be equal to some number $1-\beta$, and $\int_a^\infty f(x)dx$ must equal $\beta$.
\end{enumerate}
Figure \ref{abstract parameters} illustrates these quantities. For $\eta,\nu,\beta>0$, our formulation can be written as
\begin{subequations}\label{opt}
\begin{eqnarray}
&\underset{f}{\max}&\int_a^\infty h(x)f(x)dx\notag\\
&\text{subject to\ \ }&\int_a^\infty f(x)dx=\beta\label{c1}\\
&&f(a)=f(a+)=\eta\label{c2}\\
&&f_+'(a)\geq-\nu\label{c3}\\
&&f\text{\ convex for\ }x\geq a\label{c4}\\
&&f(x)\geq0\text{\ for\ }x\geq a\label{c5}
\end{eqnarray}
\end{subequations}
Note that we have set our objective to be $E[h(X);X\geq a]$, since $E[h(X);X<a]$ is completely known in this setting. Here $f(a+)$ denotes the right-limit at $a$, and $f(a)=f(a+)$ means that $f$ is right-continuous at $a$, implying a continuous extrapolation at $a$.

%



\subsection{Optimality Characterization}\label{sec:Opt_Char_Algo}
The solution structure of \eqref{opt} turns out to be extremely simple and is characterized by either one of two closely related cases (focusing on the region $x\geq a$). Let $\mathcal C^+[a,\infty)$ denote the class of non-negative continuous functions on $[a,\infty)$. Let
\begin{eqnarray*}
\mathcal{PL}_m^+[a,\infty)&=&\big\{f\in\mathcal C^+[a,\infty):f(x)=c_j+d_jx\text{\ for\ }x\in[y_{j-1},y_j],\ j=1,\ldots,m,{}\\
&&{}\text{\ \ where }a=y_0\leq y_1\leq\cdots\leq y_m<\infty,\ c_j,d_j\in\mathbb R,\text{\ and\ }f(x)=0\text{\ for\ }x>y_m\big\}
\end{eqnarray*}
be the set of all non-negative, continuous and piecewise linear functions on $[a,\infty)$ that have at most $m$ line segments before vanishing. We have:
\begin{theorem}
Suppose $h$ is measurable and bounded. Consider optimization \eqref{opt}. If it is feasible, then either
\begin{enumerate}
\item An optimal solution $f^*$ exists, where $f^*\in\mathcal{PL}_3^+[a,\infty)$.\label{light case}
\item An optimal solution does not exist. There exists a sequence $\{f^{(k)}\in\mathcal{PL}_3^+[a,\infty):k\geq1\}$, each $f^{(k)}$ feasible for \eqref{opt}, such that $\int_a^\infty h(x)f^{(k)}(x)dx\to Z^*$ as $k\to\infty$, where $Z^*$ is the optimal value of \eqref{opt}. Moreover, let $\{c_3^{(k)}+d_3^{(k)}x:x\in[y_2^{(k)},y_3^{(k)}]\}$ be the last line segment of $f^{(k)}$. We have $y_3^{(k)}\nearrow\infty$ and $d_3^{(k)}\searrow0$ as $k\to\infty$.\label{heavy case}
\end{enumerate}\label{main thm}
\end{theorem}

The proof of Theorem \ref{main thm} is discussed in the next sub-sections. Note that $f^*$ in the first case in Theorem \ref{main thm} is a continuous piecewise linear density, and consequently has bounded support. In the second case, as $k\to\infty$, the sequence $\{f^{(k)}:k\geq1\}$ has unboundedly increasing support endpoint ($y_3^{(k)}\nearrow\infty$), and its last line segment gets closer and more parallel to the horizontal axis ($d_3^{(k)}\searrow0$). This sequence possesses a pointwise limit, but the limit is not a valid density and has a probability mass that ``escapes" to positive infinity. 

Figures \ref{abstract light} and \ref{abtract heavy} show the tail behaviors for the two cases above. A bounded support density in the first case possesses the lightest possible tail behavior. The second case, on the other hand, can be interpreted as an extreme heavy-tail. Compare the sequence $f^{(k)}$ with a given arbitrary density. Given any fixed large enough $x$ on the real line, as $k$ grows, the decay rate of $f^{(k)}$ at the point $x$ is eventually slower than that of the given density. Since a slower decay rate is the characteristic of a fatter tail, the behavior implied by $f^{(k)}$ in a sense captures the heaviest possible tail.

\begin{centering}
\begin{minipage}{\textwidth}
  \begin{minipage}[b]{0.47\textwidth}
    \centering
    \includegraphics[scale=.24]{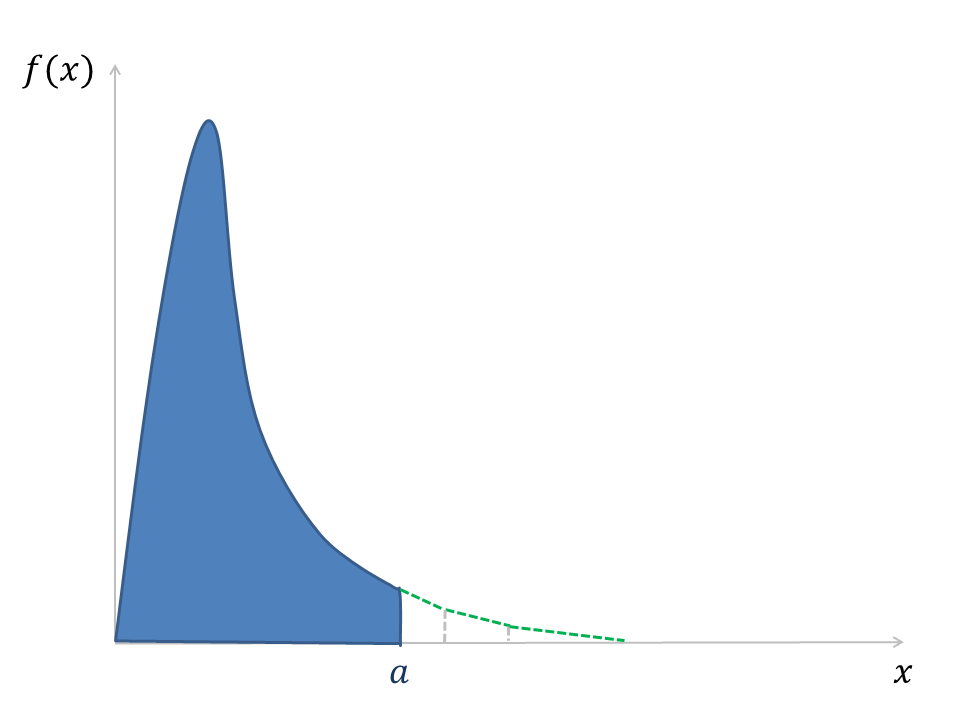}
    \captionof{figure}{Behavior of an optimal light-tailed extrapolation}
    \label{abstract light}
  \end{minipage}
  \hspace{.3cm}
  \begin{minipage}[b]{0.47\textwidth}
    \centering
    \includegraphics[scale=.31]{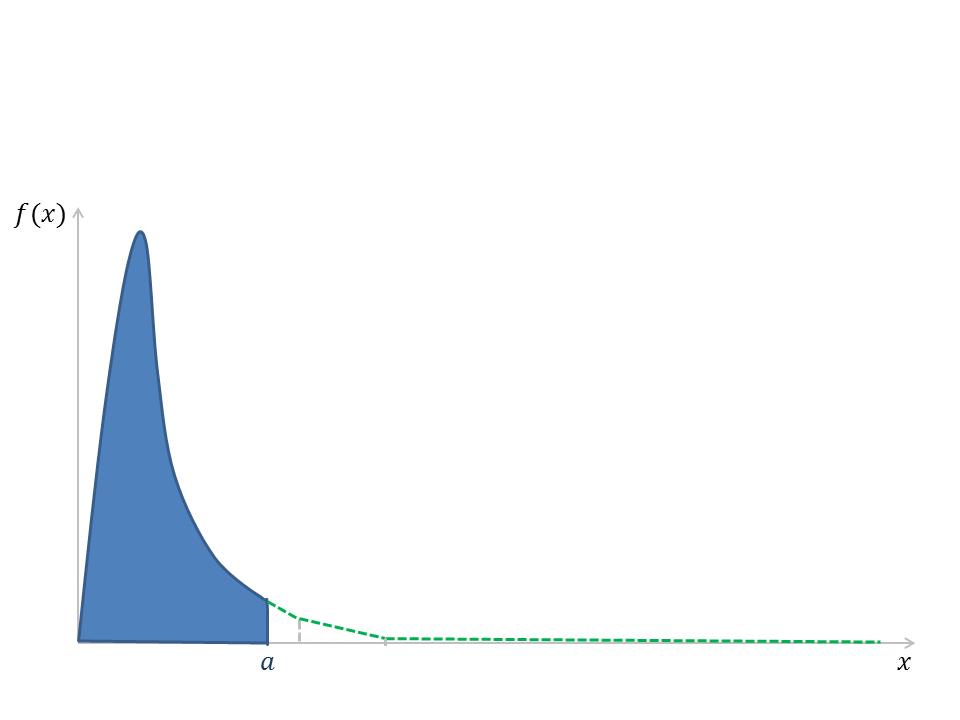}
    \captionof{figure}{Behavior of an element in an optimal heavy-tailed extrapolation sequence}
    \label{abtract heavy}
  \end{minipage}
  \end{minipage}
  \end{centering}

\subsection{Main Mathematical Developments}\label{sec:sketch}
This section presents the mathematical argument for Theorem \ref{main thm}. This development will also help construct a solution algorithm in Section \ref{sec:opt procedure}. We divide the argument into two parts. First we establish an equivalence of \eqref{opt} to a moment-constrained optimization problem under a different probability space. Second, we characterize the solution of this moment-constrained problem, which can then be converted to the solution of \eqref{opt}

We define some notations. Let $\mathbb R^+$ and $\mathbb R^-$ be the non-negative and non-positive real axis. Denote $\mathcal P(\mathcal M)$ as the set of all probability measures on a measurable space $\mathcal M$ equipped with the Borel $\sigma$-field. Let $\mathcal S_l=\{(p_1,\ldots,p_l)\in(\mathbb R^+)^l:\sum_{i=1}^lp_i=1\}$ be the $l$-dimensional probability simplex. Let $\delta(\cdot)$ be the Dirac measure. Denote $\mathcal P_n(\mathcal M)$ as the set of all finite support distributions on $\mathcal M$ with at most $n$ support points, i.e., each $\mathbb P\in\mathcal P_n(\mathcal M)$ has masses $p_1,p_2,\ldots,p_n\in\mathcal S_n$ on points $x_1,\ldots,x_n\in\mathcal M$ defined such that $\mathbb P=\sum_{i=1}^np_i\delta(x_i)$. For simplicity, since any $\mathbb P\in\mathcal P_n(\mathcal M)$ can be represented by the support points $(x_1,\ldots,x_n)\in\mathcal M^n$ (some possibly identical) and $(p_1,\ldots,p_n)\in\mathcal S_n$, we sometimes write $\mathbb P\sim(x_1,\ldots,x_n,p_1,\ldots,p_n)$ for a given $\mathbb P\in\mathcal P_n(\mathcal M)$. Moreover, we use the notation $\mathbb E[\cdot]$ to denote the associated expectation under $\mathbb P$.

For convenience, denote $\mathcal P^+=\mathcal P(\mathbb R^+)$ as the set of all probability measures concentrated on $\mathbb R^+$, and $\mathcal P_n^+=\mathcal P_n(\mathbb R^+)$ the corresponding set of measures with at most $n$ support points. The measurability of $h$ is assumed throughout the rest of the exposition.

\subsubsection{Equivalence to Moment-constrained Optimization}
We first reformulate \eqref{opt} as:
\begin{lemma}\label{thm:reformulation}
Formulation \eqref{opt} is equivalent to
\begin{subequations}\label{opt reformulated}
\begin{eqnarray}
&\underset{f}{\max}&\int_a^\infty h(x)f(x)dx\notag\\
&\text{subject to\ \ }&\int_a^\infty f(x)dx=\beta\label{cc1}\\
&&f(a)=\eta\label{cc2}\\
&&f_+'(x)\text{\ exists and is non-decreasing and right-continuous for\ }x\geq a\label{cc4}\\
&&-\nu\leq f_+'(x)\leq0\text{\ for\ }x\geq a\label{cc3}\\
&&f_+'(x)\to0\text{\ as\ }x\to\infty\label{cc5}\\
&&f(x)=\int_a^xf_+'(t)dt+\eta\text{\ for\ }x\geq a\label{cc6}
\end{eqnarray}
\end{subequations}
\end{lemma}

\proof{Proof of Lemma \ref{thm:reformulation}.}
The proof uses several elementary results from convex analysis. See Appendix \ref{sec:proofs basic} for details.\hfill\Halmos
\endproof

As a key step, we show the equivalence of \eqref{opt reformulated} to a moment-constrained program, by identifying the decision variable as $f_+'(x)$ via a one-to-one map with a probability distribution function. Let
\begin{equation}
H(x)=\int_0^x\int_0^uh(v+a)dvdu\label{H}
\end{equation}
and
\begin{equation}
\mu=\frac{\eta}{\nu}\text{\ \ and\ \ }\sigma=\frac{2\beta}{\nu}\label{def new}
\end{equation}
where $\mu,\sigma>0$ since we have assumed $\beta,\eta,\nu>0$. Our result is:
\begin{theorem}\label{thm:prob opt}
Suppose $h$ is bounded. The optimal value of \eqref{opt reformulated} is equal to that of
\begin{equation}
\begin{array}{ll}
\underset{\mathbb P}{\max}&\nu\mathbb E[H(X)]\\
\text{subject to\ \ }&\mathbb E[X]=\mu\\
&\mathbb E[X^2]=\sigma\\
&\mathbb P\in\mathcal P^+
\end{array}\label{prob opt main}
\end{equation}
Here the decision variable is a probability measure $\mathbb P\in\mathcal P^+$, and $\mathbb E[\cdot]$ is the corresponding expectation. Moreover, there is a one-to-one correspondence between the feasible solutions to \eqref{opt reformulated} and \eqref{prob opt main}, given by $f_+'(x+a)=\nu(p(x)-1)$ for $x\in\mathbb R^+$, where $f_+'$ is the right derivative of a feasible solution $f$ of \eqref{opt reformulated} such that $f(x)=\int_a^xf_+'(t)dt+\eta$ for $x\geq a$, and $p$ is a probability distribution function that is associated with a feasible probability measure over $\mathbb R^+$ in \eqref{prob opt main}.
\label{main reduction}
\end{theorem}
\proof{Proof of Theorem \ref{main reduction}.}
The key step of the proof uses integration by parts and an explicit construction of a linear transformation between $f_+'$ and a probability distribution function $p$. See Appendix \ref{sec:proofs basic} for details.\hfill\Halmos
\endproof

Note that $\nu$ appears in the objective function in \eqref{prob opt main} whose optimal value matches that of program \eqref{opt reformulated}.

\subsubsection{Further Reduction and Optimality Characterization}
Next we characterize the optimality structure for \eqref{prob opt main}, a generalized moment problem in the form of an infinite-dimensional linear program. Using existing terminology, we call an optimization program \emph{consistent} if there exists a feasible solution, and \emph{solvable} if there exists an optimal solution.


For convenience, denote $OPT(\mathcal D)$ as the program
\begin{equation*}
\begin{array}{ll}
\underset{\mathbb P}{\max}&\nu\mathbb E[H(X)]\\
\text{subject to\ \ }&\mathbb E[X]=\mu\\
&\mathbb E[X^2]=\sigma\\
&\mathbb P\in\mathcal D
\end{array}
\end{equation*}
where $H,\mu,\sigma$ are defined in \eqref{H} and \eqref{def new}, and $\mathcal D$ is a collection of probability measures on $\mathbb R$. For example, program \eqref{prob opt main} is denoted as $OPT(\mathcal P^+)$. Moreover, let $Z(\mathbb P)=\nu\mathbb E[H(X)]$ be the objective function of $OPT(\mathcal D)$ in terms of $\mathbb P$. We have:

\begin{theorem}
Program \eqref{prob opt main}, or equivalently $OPT(\mathcal P^+)$, has the same optimal value as $OPT(\mathcal P_3^+)$. \label{equivalence finite}
\end{theorem}

\proof{Proof of Theorem \ref{equivalence finite}.}
Follows from a classical result on the extreme points of moment sets. See Appendix \ref{sec:proofs basic}.\hfill\Halmos
\endproof

Next we derive some properties regarding the optimality of $OPT(\mathcal P_3^+)$:
\begin{proposition}
Consider $OPT(P_3^+)$ that is consistent. The optimal value $Z^*$ is either achieved at some $\mathbb P^*\in\mathcal P_3^+$, or there exists a sequence of feasible $\mathbb P^{(k)}\in\mathcal P_3^+$ such that $Z(\mathbb P^{(k)})\to Z^*$. In the second case, each $\mathbb P^{(k)}\sim(x_1^{(k)},x_2^{(k)},x_3^{(k)},p_1^{(k)},p_2^{(k)},p_3^{(k)})$, such that either $(x_1^{(k)},x_2^{(k)},x_3^{(k)},p_1^{(k)},p_2^{(k)},p_3^{(k)})\to(x_1^*,x_2^*,\infty,p_1^*,p_2^*,0)$ for some $x_1^*,x_2^*\in\mathbb R^+$ and $(p_1^*,p_2^*)\in\mathcal S_2$ (where $x_1^*$ and $x_2^*$ are possibly identical), or $(x_1^{(k)},x_2^{(k)},x_3^{(k)},p_1^{(k)},p_2^{(k)},p_3^{(k)})\to(x_1^*,\infty,\infty,1,0,0)$ for some $x_1^*\in\mathbb R^+$.
\label{char}
\end{proposition}

\proof{Proof of Proposition \ref{char}.}
See Appendix \ref{sec:proofs basic}.\hfill\Halmos
\endproof

We are now ready to show Theorem \ref{main thm}:
\proof{Proof of Theorem \ref{main thm}.}
Convert the original optimization \eqref{opt} into \eqref{prob opt main} by Lemma \ref{thm:reformulation} and Theorem \ref{main reduction}. If \eqref{prob opt main} is consistent, then, by Theorem \ref{equivalence finite}, its optimal value is attained by the two cases in Proposition \ref{char}. Note that that any solution $\mathbb P\in\mathcal P_3[0,\infty)$ represented by $(x_1,x_2,x_3,p_1,p_2,p_3)$ (where some of $x_1$, $x_2$ and $x_3$ are possibly identical) admits one-to-one correspondence with a solution $f$ in \eqref{opt}, via $f_+'(x+a)=\nu(p(x)-1)$ in Theorem \ref{main reduction}, giving
$$f_+'(x)=\left\{\begin{array}{ll}
-\nu&\text{\ for\ }a\leq x<x_1+a\\
-\nu(1-p_1)&\text{\ for\ }x_1+a\leq x<x_2+a\\
-\nu(1-p_1-p_2)&\text{\ for\ }x_2+a\leq x<x_3+a\\
0&\text{\ for\ }x_3+a\leq x
\end{array}\right.$$
and hence
\begin{equation}
f(x)=\left\{\begin{array}{ll}
\eta-\nu(x-a)&\text{\ for\ }a\leq x\leq x_1+a\\
\eta-\nu x_1-\nu(1-p_1)(x-a-x_1)&\text{\ for\ }x_1+a\leq x\leq x_2+a\\
\eta-\nu x_1-\nu(1-p_1)(x_2-x_1)-\nu(1-p_1-p_2)(x-a-x_2)&\text{\ for\ }x_2+a\leq x\leq x_3+a\\
0&\text{\ for\ }x_3+a\leq x
\end{array}\right.\label{interim4new}
\end{equation}
The first case in Proposition \ref{char} thus concludes Part \ref{light case} of Theorem \ref{main thm}. In the second case in Proposition \ref{char}, $x_3^{(k)}\to\infty$ and $p_3^{(k)}\to0$ so that $1-p_1^{(k)}-p_2^{(k)}\to0$. Using \eqref{interim4new}, we conclude Part \ref{heavy case} of Theorem \ref{main thm}.\hfill\Halmos
\endproof

We close this section with two results. First is on the consistency of programs \eqref{opt} and \eqref{prob opt main}:
\begin{lemma}
Program \eqref{prob opt main} is consistent if and only if $\sigma\geq\mu^2$. Correspondingly, program \eqref{opt} is consistent if and only if $\eta^2\leq2\beta\nu$. When $\sigma=\mu^2$, \eqref{prob opt main} has only one feasible solution given by $\delta(\mu)$. Correspondingly, when $\eta^2=2\beta\nu$, \eqref{opt} has only one feasible solution given by $f(x)=\eta-\nu(x-a)$ for $x\geq a$. \label{consistency}
\end{lemma}
\proof{Proof of Lemma \ref{consistency}.}
See Appendix \ref{proofs opt procedure}.\hfill\Halmos
\endproof

Graphically, $\eta^2>2\beta\nu$ implies that $\beta$ is smaller than the area under the straight line starting from the point $(a,\eta)$ down to the $x$-axis with slope $-\nu$. Hence no convex extrapolation can be drawn under this condition.

Next, we show that the boundedness assumption on $h$ is nearly essential, in the sense that any polynomially growing $h$ leads to an infinite optimal value for \eqref{opt}:
\begin{proposition}
Suppose $\eta^2<2\beta\nu$ and $h(x)=\Omega(x^\epsilon)$ as $x\to\infty$ for some $\epsilon>0$. The optimal value of \eqref{opt} is $\infty$.
\label{counter example}
\end{proposition}

\proof{Proof of Proposition \ref{counter example}.}
The proof explicitly constructs a sequence of feasible solutions that lead to exploding objective values. See Appendix \ref{sec:proofs basic}.\hfill\Halmos
\endproof


\section{Optimization Procedure for Quasi-concave Objectives}\label{sec:opt procedure}
This section develops a numerical solution algorithm for our worst-case optimization presented in Section \ref{sec:basic}. In building our algorithm, we focus on $h$ that satisfies the following stronger assumption, which covers many natural scenarios including the two examples in the Introduction.
\begin{assumption}
The function $h:\mathbb R\to\mathbb R^+$ is bounded, and is non-decreasing in $[a,c)$ and non-increasing in $(c,\infty)$ for some constant $a\leq c\leq\infty$ (i.e. $c$ can possibly be $\infty$).\label{regularity}
\end{assumption}

Assumption \ref{regularity} implies that $h$ is quasi-concave. The non-negativity of $h$ is assumed without loss of generality when applied to optimization \eqref{opt}. Because $h$ is bounded, one can always add a sufficiently large constant, say $C$, to make $h$ non-negative. Note that we have $E[h(X);X\geq a]=E[h(X)+C;X\geq a]-CP(X\geq a)=E[h(X)+C;X\geq a]-C\beta$, and so one can solve $E[h(X)+C;X\geq a]$ and recover $E[h(X);X\geq a]$.

We impose an additional mild regularity assumption:
\begin{assumption}
The limit
\begin{equation}
\lambda=\lim_{x\to\infty}\frac{H(x)}{x^2}\label{definitions}
\end{equation}
where $H$ is defined in \eqref{H}, exists and is finite.\label{H limit}
\end{assumption}
Note that when $h$ is bounded, $H(x)=O(x^2)$ as $x\to\infty$, and $\limsup_{x\to\infty}H(x)/x^2<\infty$. The essence of Assumption \ref{H limit} is on the existence of the limit.

Under Assumption \ref{H limit}, denote
\begin{equation}
  W(x_1)=\nu\left(\frac{\sigma-\mu^2}{\sigma-2\mu x_1+x_1^2}H(x_1)+\frac{(\mu-x_1)^2}{\sigma-2\mu x_1+x_1^2}H\left(\frac{\sigma-\mu x_1}{\mu-x_1}\right)\right)\label{line search1}
  \end{equation}
for $x_1\in[0,\mu)$ and $W(\mu):=\nu(H(\mu)+\lambda(\sigma-\mu^2))$, where $\mu$ and $\sigma$ are defined in \eqref{def new}. 
We have the following strengthened version of Theorem \ref{main thm}:
\begin{theorem}
Under Assumption \ref{regularity},
\begin{enumerate}
\item The conclusions of Theorem \ref{main thm} hold with $\mathcal{PL}_3^+[a,\infty)$ replaced by $\mathcal{PL}_2^+[a,\infty)$.\label{p1}
\item Suppose $\eta^2<2\beta\nu$ and Assumption \ref{H limit} holds additionally. The optimal value of \eqref{opt} is given by $\max_{x_1\in[0,\mu]}W(x_1)$.\label{pt2}
\item Suppose $\eta^2<2\beta\nu$ and Assumption \ref{H limit} holds additionally. If $\text{argmax}_{x_1\in[0,\mu]}W(x_1)\cap[0,\mu)\neq\emptyset$, then an optimal solution to \eqref{opt} is given by
$$f^*(x)=\left\{\begin{array}{ll}
\eta-\nu(x-a)&\text{\ for\ }a\leq x\leq x_1^*+a\\
\eta-\nu x_1^*-\nu\frac{(\mu-x_1^*)^2}{\sigma-2\mu x_1^*+{x_1^*}^2}(x-a-x_1^*)&\text{\ for\ }x_1^*+a\leq x\leq\frac{\sigma-\mu x_1^*}{\mu-x_1^*}+a\\
0&\text{\ for\ }\frac{\sigma-\mu x_1^*}{\mu-x_1^*}+a\leq x
\end{array}\right.$$
where $x_1^*\in\text{argmax}_{x_1\in[0,\mu]}W(x_1)\cap[0,\mu)$. Otherwise, we have $\text{argmax}_{x_1\in[0,\mu]}W(x_1)=\{\mu\}$, and there exists a sequence of feasible solutions $f^{(k)}$ with $\int_a^\infty h(x)f^{(k)}(x)dx\to Z^*$, where $Z^*$ is the optimal value of \eqref{opt}. $f^{(k)}\to f^*$ pointwise where
$$f^*(x)=\left\{\begin{array}{ll}\eta-\nu(x-a)&\text{\ for\ }a\leq x\leq\mu+a\\
0&\text{\ for\ }\mu+a\leq x
\end{array}\right.$$
The second case can occur only when $\lambda>0$.\label{pt3}
\end{enumerate}
\label{main thm simplified}
\end{theorem}

Part \ref{p1} of Theorem \ref{main thm simplified} simplifies the search space of densities in \eqref{opt} from three to two linear segments. Because of this simplification, solving \eqref{opt} reduces to finding the first kink of the optimal density (or sequence of densities), equivalently the first support point of the reformulation \eqref{prob opt main}. This can be done by a one-dimensional line search $\max_{x_1\in[0,\mu]}W(x_1)$ in Part \ref{pt2} of the theorem. 

Part \ref{pt3} of Theorem \ref{main thm simplified} describes how to distinguish between the light- and heavy-tail cases in Theorem \ref{main thm} by looking at the location of $x_1^*$. The former case occurs when there exists a $x_1^*$ in $[0,\mu)$, and the latter occurs otherwise. Note that $f^*(x)=0,\ x\geq\mu+a$ in the pointwise limit of $f^{(k)}$ in Part \ref{pt3} of Theorem \ref{main thm simplified} is a consequence of the last line segment of $f^{(k)}$ getting increasingly closer and more parallel to the $x$-axis.

Algorithm 1 summarizes the procedure for obtaining the optimal value of \eqref{opt}.

\noindent\hrulefill
\vspace{-5mm}

\noindent\textbf{Algorithm 1: Procedure for finding the optimal value of \eqref{opt}}
\vspace{-7mm}

\noindent\hrulefill
\vspace{-4mm}

\noindent\textbf{Inputs: }

\begin{enumerate}
\item The function $h$ that satisfies Assumptions \ref{regularity} and \ref{H limit}.

\item The parameters $\beta,\eta,\nu>0$.
\end{enumerate}

\noindent\textbf{Procedure: }

\begin{enumerate}
\item If $\eta^2>2\beta\nu$, there is no feasible solution.
\item If $\eta^2=2\beta\nu$, the optimal value is $\nu H(\mu)$.
\item If $\eta^2<2\beta\nu$, the optimal value is given by $\max_{x_1\in[0,\mu]}W(x_1)$.
\end{enumerate}

\noindent\hrulefill

\bigskip

The rest of this section provides the developments for proving Theorem \ref{main thm simplified}. First we introduce the following condition:
\begin{assumption}
$H$ is convex and $H'$ satisfies a convex-concave property, i.e. $H'(x)$ is convex for $x\in(0,c)$ and concave for $x\in(c,\infty)$, for some $0\leq c\leq\infty$.\label{regularity H}
\end{assumption}

With Assumption \ref{regularity H}, Theorem \ref{equivalence finite} can be strengthened to:
\begin{proposition}
Under Assumption \ref{regularity H}, $OPT(\mathcal P^+)$ has the same optimal value as $OPT(\mathcal P_2^+)$.\label{2-supp}
\end{proposition}

\proof{Proof of Proposition \ref{2-supp}.}
See Appendix \ref{proofs opt procedure}.\hfill\Halmos
\endproof

This allows us to focus on one of the support points of $OPT(\mathcal P_2^+)$ in the solution scheme, leading to the following proposition:
\begin{proposition}
Under Assumptions \ref{H limit} and \ref{regularity H}, consider $OPT(\mathcal P_2^+)$ with $\sigma>\mu^2$ and let $Z^*$ be its optimal value.
\begin{enumerate}
\item If there exists an optimal solution in $\mathcal P_2^+$, then this solution has distinct support points and is represented by $(x_1^*,x_2^*,p_1^*,p_2^*)$, where $x_1^*\in\text{argmax}_{x_1\in[0,\mu)}W(x_1)$ and
    \begin{equation}
    x_2^*=\frac{\sigma-\mu x_1^*}{\mu-x_1^*},\ \ \ \ p_1^*=\frac{\sigma-\mu^2}{\sigma-2\mu x_1^*+{x_1^*}^2},\ \ \ \ p_2^*=\frac{(\mu-x_1^*)^2}{\sigma-2\mu x_1^*+{x_1^*}^2}\label{opt sol}
    \end{equation}
Moreover, $Z^*=\max_{x_1\in[0,\mu)}W(x_1)$.\label{case1}
\item If there does not exist an optimal solution, then there must exist a sequence $\mathbb P^{(k)}\sim(x_1^{(k)},x_2^{(k)},p_1^{(k)},p_2^{(k)})\to(\mu,\infty,1,0)$. Moreover, $Z^*=\nu(H(\mu)+\lambda(\sigma-\mu^2))$.\label{case2}
\item $Z^*=\max_{x_1\in[0,\mu]}W(x_1)$\label{case overall}
\end{enumerate}
\label{char1}
\end{proposition}

\proof{Proof of Proposition \ref{char1}.}
See Appendix \ref{proofs opt procedure}.\hfill\Halmos
\endproof

The following corollary provides a simple sufficient conditions for guaranteeing the light-tail case in the solution scheme:
\begin{corollary}
Suppose Assumptions \ref{regularity} and \ref{H limit} hold and \eqref{opt} is consistent. An optimal solution for \eqref{opt} must exist if $\lambda=0$.\label{special}
\end{corollary}

\proof{Proof of Corollary \ref{special}.}
By Lemma \ref{consistency}, consistency of \eqref{opt} implies $\sigma\geq\mu^2$. By Theorem \ref{thm:prob opt} and Proposition \ref{2-supp}, it suffices to consider the equivalent program $OPT(\mathcal P_2^+)$. Suppose $\lambda=0$. If $\sigma=\mu^2$, then $\delta(\mu)$ is an optimal solution. If $\sigma>\mu^2$, then by Proposition \ref{char1}, if there is no optimal solution, its optimal value must be $\nu(H(\mu)+\lambda(\sigma-\mu^2))=\nu H(\mu)$, which is attained by $\delta(\mu)$ and leads to a contradiction (to both the hypotheses of no optimal solution and $\sigma>\mu^2$).\hfill\Halmos
\endproof

We are now ready to show Theorem \ref{main thm simplified}:
\proof{Proof of Theorem \ref{main thm simplified}.}
\underline{Proof of \ref{p1}.} Assumption \ref{regularity} implies Assumption \ref{regularity H}. By Theorem \ref{thm:prob opt} and Proposition \ref{2-supp}, program \eqref{opt} has the same optimal value as that of $OPT(\mathcal P_2^+)$. Similar to the proof of Theorem \ref{main thm}, the result follows by noting that any $\mathbb P\in\mathcal P_2^+$ represented by $(x_1,x_2,p_1,p_2)$ (with possibly identical $x_i$'s) admits one-to-one correspondence with a solution $f$ in \eqref{opt}, via $f_+'(x+a)=\nu(p(x)-1)$ in Theorem \ref{main reduction}, giving
$$f_+'(x)=\left\{\begin{array}{ll}
-\nu&\text{\ for\ }a\leq x<x_1+a\\
-\nu p_2&\text{\ for\ }x_1+a\leq x<x_2+a\\
0&\text{\ for\ }x_2+a\leq x
\end{array}\right.$$
and hence
\begin{equation}
f(x)=\left\{\begin{array}{ll}
\eta-\nu(x-a)&\text{\ for\ }a\leq x\leq x_1+a\\
\eta-\nu x_1-\nu p_2(x-a-x_1)&\text{\ for\ }x_1+a\leq x\leq x_2+a\\
0&\text{\ for\ }x_2+a\leq x
\end{array}\right.\label{solution form}
\end{equation}

\underline{Proof of \ref{pt2}.} The condition $\eta^2<2\beta\nu$ is equivalent to $\sigma>\mu^2$. The conclusion follows from Part \ref{case overall} in Proposition \ref{char1}.

\underline{Proof of \ref{pt3}.} The first case is obtained by substituting $x_1^*\in\text{argmax}_{x_1\in[0,\mu)}W(x_1)$ and $x_2^*,p_2^*$ from \eqref{opt sol}, in Part \ref{case1} in Proposition \ref{char1}, into \eqref{solution form}. The second case is obtained by substituting $(x_1^{(k)},x_2^{(k)},p_1^{(k)},p_2^{(k)})$ in Part \ref{case2} in Proposition \ref{char1} into \eqref{solution form} and taking the limit. The last conclusion follows from Corollary \ref{special}.\hfill\Halmos
\endproof


\section{Formulation and Procedure under Data-driven Environment}\label{sec:numerics data}
Sections \ref{sec:basic} and \ref{sec:opt procedure} have discussed our worst-case approach in the abstract setting where the values of the needed parameters $\beta,\eta,\nu$ are completely known. In practice, these parameters are not directly specified. Instead, they are calibrated from data in the non-tail region.
Suppose we obtain confidence intervals (CIs) for $P(X>a)$ and $f(a)$ and a lower confidence bound for $f_-'(a)$, jointly with confidence level $1-\alpha$. Denote them as $[\underline\beta,\overline\beta]$, $[\underline\eta,\overline\eta]$ and $-\overline\nu$. Suppose $\underline\beta,\overline\beta,\underline\eta,\overline\eta,\overline\nu>0$. We substitute these estimates for the exact values of $\beta$, $\eta$ and $-\nu$ in our worst-case bound for $E[h(X);X\geq a]$:
\begin{equation}
\begin{array}{ll}
\underset{f}{\max}&\int_a^\infty h(x)f(x)dx\\
\text{subject to\ \ }&\underline\beta\leq\int_a^\infty f(x)dx\leq\overline\beta\\
&\underline\eta\leq f(a)=f(a+)\leq\overline\eta\\
&f_+'(a)\geq-\overline\nu\\
&f(x)\text{\ convex for\ }x\geq a\\
&f(x)\geq0\text{\ for\ }x\geq a
\end{array}\label{opt relaxed}
\end{equation}
It is immediate that the optimal value of \eqref{opt relaxed} carries the following statistical guarantee:
\begin{proposition}
Suppose that $[\underline\beta,\overline\beta]$, $[\underline\eta,\overline\eta]$ and $-\overline\nu$ are the joint $(1-\alpha)$-level CIs for $P(X>a)$ and $f(a)$, and lower confidence bound for $f_-'(a)$. Then with probability $1-\alpha$ (with respect to the data) optimization \eqref{opt relaxed} gives an upper bound for $E[h(X);X\geq a]$ under the assumption that $f(x)$ is convex for $x\geq a$ and $f(a)=f(a+)$.\label{stat}
\end{proposition}

\proof{Proof of Proposition \ref{stat}.}
Let $f_{true}(x),\ x\geq a$ be the ground-true density, and $Z_{true}=\int_a^\infty h(x)f_{true}(x)dx$. Let $Z^*$ and $\mathcal F$ be the optimal value and feasible region of \eqref{opt relaxed}. If $f_{true}\in\mathcal F$, then $Z^*\geq Z_{true}$. Hence $P_{\text{data}}(Z^*\geq Z_{true})\geq P_{\text{data}}(f_{true}\in\mathcal F)=1-\alpha$, where $P_{\text{data}}$ denotes the probability with respect to the data.\hfill\Halmos
\endproof

For $h$ that has support spanning across both $X<a$ and $X\geq a$, one approach is to estimate $E[h(X);X<a]$ separately from the computation of the worst-case bound from \eqref{opt relaxed}. The former can be done typically by using the empirical mean as the non-tail region $X<a$ possesses more data to rely on. This segregated approach, however, only allows the conditions of valid probability density on the whole real line (e.g., $\int_{\mathbb R}f(x)dx=1$) and the continuity at $a$ to hold approximately but not exactly. 


The following result presents the optimality structure for \eqref{opt relaxed} in parallel to formulation \eqref{opt}.

\begin{theorem}
Suppose $h$ is bounded. Consider optimization \eqref{opt relaxed}. If it is feasible, then either
\begin{enumerate}
\item An optimal solution $f^*$ exists, where $f^*\in\mathcal{PL}_3^+[a,\infty)$.
\item An optimal solution does not exist. There exists a sequence $\{f^{(k)}\in\mathcal{PL}_3^+[a,\infty):k\geq1\}$, each $f^{(k)}$ feasible for \eqref{opt}, such that $\int_a^\infty h(x)f^{(k)}(x)dx\to Z^*$ as $k\to\infty$, where $Z^*$ is the optimal value of \eqref{opt relaxed}. Moreover, let $\{c_3^{(k)}+d_3^{(k)}x:x\in[y_2^{(k)},y_3^{(k)}]\}$ be the last line segment of $f^{(k)}$. We have $y_3^{(k)}\nearrow\infty$ and $d_3^{(k)}\searrow0$ as $k\to\infty$.
\end{enumerate}\label{main thm relaxed}
\end{theorem}
\proof{Proof of Theorem \ref{main thm relaxed}.}
See Appendix \ref{sec:data driven proofs}.\hfill\Halmos
\endproof


%
Define
\begin{equation}
\underline\mu=\frac{\underline\eta}{\overline\nu},\ \overline\mu=\frac{\overline\eta}{\overline\nu},\ \underline\sigma=\frac{2\underline\beta}{\overline\nu},\ \overline\sigma=\frac{2\overline\beta}{\overline\nu}\label{definitions1}
\end{equation}
where $\underline\mu,\overline\mu,\underline\sigma,\overline\sigma>0$ since we have assumed $\underline\beta,\overline\beta,\underline\eta,\overline\eta,\overline\nu>0$. Define
$$\mathcal W(x,\omega,\rho)=\overline\nu\left(\frac{\rho-\omega^2}{\rho-2\omega x+x^2}H(x)+\frac{(\omega-x)^2}{\rho-2\omega x+x^2}H\left(\frac{\rho-\omega x}{\omega-x}\right)\right)$$
with $\mathcal W(\omega,\omega,\rho):=\overline\nu(H(\omega)+\lambda(\rho-\omega^2))$, where $H$ and $\lambda$ are defined as in \eqref{H} and \eqref{definitions}.

For convenience, we also denote
\begin{align*}
&\mathcal K(x;x_1,\omega,\rho)=\left\{\begin{array}{ll}
\overline\nu\omega-\overline\nu(x-a)&\text{\ for\ }a\leq x\leq x_1+a\\
\overline\nu\omega-\overline\nu x_1-\overline\nu\frac{(\omega-x_1)^2}{\rho-2\omega x_1+x_1^2}(x-a-x_1)&\text{\ for\ }x_1+a\leq x\leq\frac{\rho-\omega x_1}{\omega-x_1}+a\\
0&\text{\ for\ }\frac{\rho-\omega x_1}{\omega-x_1}+a\leq x
\end{array}\right.
\end{align*}

Our data-integrated optimization \eqref{opt relaxed} possesses the following consistency property in parallel to the fixed-parameter case in Lemma \ref{consistency}:
\begin{lemma}
Program \eqref{opt relaxed} is consistent if and only if $\underline\eta^2\leq2\overline\beta\overline\nu$ or equivalently $\overline\sigma\geq\underline\mu^2$. When $\underline\eta^2=2\overline\beta\overline\nu$ or equivalently $\overline\sigma=\underline\mu^2$, \eqref{opt relaxed} has only one feasible solution given by $f(x)=\underline\eta-\overline\nu(x-a)$ for $x\geq a$. \label{consistency relaxed}
\end{lemma}
\proof{Proof of Lemma \ref{consistency relaxed}.}
The proof is similar to Lemma \ref{consistency} and hence skipped.\hfill\Halmos
\endproof

The following provides the solution scheme for our data-integrated optimization \eqref{opt relaxed}:
\begin{theorem}
Under Assumption \ref{regularity},
\begin{enumerate}
\item The conclusions of Theorem \ref{main thm relaxed} hold with $\mathcal{PL}_3^+[a,\infty)$ replaced by $\mathcal{PL}_2^+[a,\infty)$.\label{p1}
\item Suppose $\underline\eta^2<2\overline\beta\overline\nu$ and Assumption \ref{H limit} holds additionally. The optimal value of \eqref{opt relaxed} is given by
\begin{equation}
\max\left\{\max_{\rho\in[\underline\sigma\vee \overline\mu^2,\overline\sigma],x_1\in[0,\overline\mu]}\mathcal W(x_1,\overline\mu,\rho),\max_{\omega\in[\underline\mu,\overline\mu\wedge\sqrt{\overline\sigma}],x_1\in[0,\omega]}\mathcal W(x_1,\omega,\overline\sigma)\right\}\label{new opt}
\end{equation}
\item Suppose $\underline\eta^2<2\overline\beta\overline\nu$ and Assumption \ref{H limit} holds additionally. Suppose $\max_{\rho\in[\underline\sigma\vee \overline\mu^2,\overline\sigma],x_1\in[0,\overline\mu]}\mathcal W(x_1,\overline\mu,\rho)\geq\max_{\omega\in[\underline\mu,\overline\mu\wedge\sqrt{\overline\sigma}],x_1\in[0,\omega]}\mathcal W(x_1,\omega,\overline\sigma)$. If there exists $(\rho^*,x_1^*)\in\text{argmax}_{\rho\in[\underline\sigma\vee \overline\mu^2,\overline\sigma],x_1\in[0,\overline\mu]}\mathcal W(x_1,\overline\mu,\rho)$ such that $x_1^*\in[0,\overline\mu)$, then an optimal solution to \eqref{opt relaxed} is given by $f^*(x)=\mathcal K(x;x_1^*,\overline\mu,\rho^*)$. Otherwise, there exists a sequence of feasible solutions $f^{(k)}$ with $\int_a^\infty h(x)f^{(k)}(x)dx\to Z^*$, the optimal value of \eqref{opt relaxed}, such that $f^{(k)}\to f^*$ pointwise where
$$f^*(x)=\left\{\begin{array}{ll}\overline\eta-\overline\nu(x-a)&\text{\ for\ }a\leq x\leq\overline\mu+a\\
0&\text{\ for\ }\overline\mu+a\leq x
\end{array}\right.$$
which can occur only when $\lambda>0$. On the other hand, suppose $\max_{\rho\in[\underline\sigma\vee \overline\mu^2,\overline\sigma],x_1\in[0,\overline\mu]}\mathcal W(x_1,\overline\mu,\rho)<\max_{\omega\in[\underline\mu,\overline\mu\wedge\sqrt{\overline\sigma}],x_1\in[0,\omega]}\mathcal W(x_1,\omega,\overline\sigma)$. If there exists $(\omega^*,x_1^*)\in\text{argmax}_{\omega\in[\underline\mu,\overline\mu\wedge\sqrt{\overline\sigma}],x_1\in[0,\omega]}\mathcal W(x_1,\omega,\overline\sigma)$ such that $x_1^*\in[0,\omega^*)$, then an optimal solution to \eqref{opt relaxed} is given by $f^*(x)=\mathcal K(x;x_1^*,\omega^*,\overline\sigma)$. Otherwise, there exists a sequence of feasible solutions $f^{(k)}$ with $\int_a^\infty h(x)f^{(k)}(x)dx\to Z^*$, such that $f^{(k)}\to f^*$ pointwise where
$$f^*(x)=\left\{\begin{array}{ll}\overline\nu\omega^*-\overline\nu(x-a)&\text{\ for\ }a\leq x\leq\omega^*+a\\
0&\text{\ for\ }\omega^*+a\leq x
\end{array}\right.$$
which again can occur only when $\lambda>0$.
\end{enumerate}
\label{main thm simplified relaxed}
\end{theorem}
\proof{Proof of Theorem \ref{main thm simplified relaxed}.}
Optimization \eqref{new opt} follows from a reduction of the inequality-based generalized moment problem converted from \eqref{opt relaxed} into two subproblems. Appendix \ref{sec:data driven proofs} provides the constituent propositions and further details.\hfill\Halmos
\endproof

Note that in the current setting it is less straightforward to transform a problem with a general bounded $h$ into one that has a non-negative $h$ than in Section \ref{sec:opt procedure} (see the discussion after Assumption \ref{regularity}), since the probability mass assigned to $[a,\infty)$ is now bounded between $\underline\beta$ and $\overline\beta$ instead of being a single specified value.

Algorithm 2 presents our procedure for solving \eqref{opt relaxed}.

\noindent\hrulefill
\vspace{-5mm}

\noindent\textbf{Algorithm 2: Procedure for Finding the Optimal Value of \eqref{opt relaxed}}
\vspace{-7mm}

\noindent\hrulefill
\vspace{-4mm}

\noindent\textbf{Inputs: }

\begin{enumerate}
\item The function $h$ that satisfies Assumptions \ref{regularity} and \ref{H limit}. 

\item The parameters $\underline\beta,\overline\beta,\underline\eta,\overline\eta,\overline\nu>0$.
\end{enumerate}

\noindent\textbf{Procedure: }

\begin{enumerate}
\item If $\underline\eta^2>2\overline\beta\overline\nu$, there is no feasible solution.
\item If $\underline\eta^2=2\overline\beta\overline\nu$, the optimal value is $\overline\nu H(\underline\mu)$.
\item If $\underline\eta^2<2\overline\beta\overline\nu$, the optimal value is
$$\max\left\{\max_{\rho\in[\underline\sigma\vee \overline\mu^2,\overline\sigma],x_1\in[0,\overline\mu]}\mathcal W(x_1,\overline\mu,\rho),\max_{\omega\in[\underline\mu,\overline\mu\wedge\sqrt{\overline\sigma}],x_1\in[0,\omega]}\mathcal W(x_1,\omega,\overline\sigma)\right\}$$
\end{enumerate}

\noindent\hrulefill


\section{Numerical Examples}\label{sec:numerics}
We present some numerical performance of our algorithm. We first consider several elementary examples, and then we will revisit the two examples in the Introduction.

\subsection{Elementary Examples}
We consider three examples to demonstrate Algorithm 1.

\begin{figure}[h]
\centering
\includegraphics[scale = 0.75]{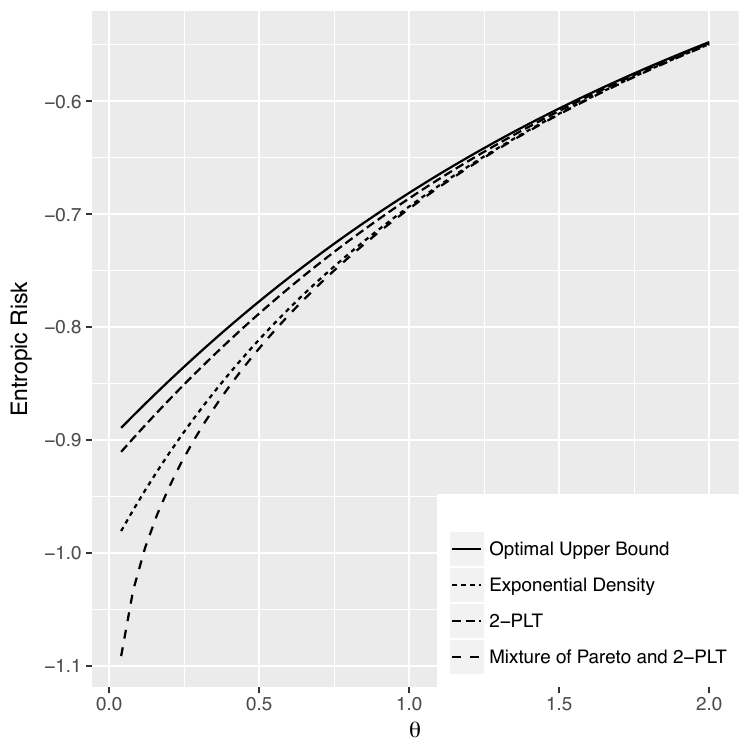}
\caption{\small{Optimal upper bound and comparison with parametric extrapolations for the entropic risk measure.}}
\label{fig:entropic_risk}
\end{figure}

\noindent\textbf{Entropic Risk Measure: } The entropic risk measure (e.g., \cite{follmer2011stochastic}) captures the risk aversion of users through the exponential utility function. It is defined as
\begin{equation}
\rho(X) = \frac{1}{\theta} \log\left(E\left[e^{-\theta X}\right]\right)\label{entropic risk}
\end{equation}
where $\theta > 0$ is the parameter of risk aversion. In the case when the distribution of the random variable $X$ is known only up to some point $a$, we can find the worst case value of the entropic risk measure subject to tail uncertainty by solving the optimization problem
\begin{equation}
\max_{P\in\mathcal A} \frac{1}{\theta} \log\left(E\left[e^{-\theta X}\right]\right) = \frac{1}{\theta} \log\left(E[e^{-\theta X};X\leq a]  + \max_{P\in\mathcal A} E\left[e^{-\theta X};X> a\right]\right)\label{robust entropic risk}
\end{equation}
where $\mathcal A$ denotes the set of convex tails that match the given non-tail region. Since the function $e^{-\theta X}$ satisfies Assumptions \ref{regularity} and \ref{H limit}, we can apply Algorithm 1 to the second term of the RHS of \eqref{robust entropic risk}. The thick line in Figure \ref{fig:entropic_risk} represents the worst-case value of the entropic risk measure for different values of the parameter $\theta$ in the case when $X$ is known to have a standard exponential distribution $Exp(1)$ up to $a= -\log(0.7)$ ($i.e.$ $a$ is the $70$-percentile and $\beta = \eta = \nu = 0.7$). For comparison, we also calculate and plot the entropic risk measure for several fitted probability distributions: $Exp(1)$, two-segment continuous piecewise linear tail denoted as 2-PLT (two such instances in Figure \ref{fig:entropic_risk}), and mixtures of 2-PLT and shifted Pareto. Clearly, the worst-case values bound those calculated from the candidate parametric models, with the gap diminishing as $\theta$ increases.
\\
 %

\noindent\textbf{The Newsvendor Problem: } The classical newsvendor problem maximizes the profit of selling a perishable product by fulfilling demand using a stock level decision, i.e.,
\begin{equation}
\max_qE[p\min(q,D)]-cq\label{newsvendor}
\end{equation}
where $D$ is the demand random variable, $p$ and $c$ are the selling and purchase prices per product, and $q$ is the stock quantity to be determined. We assume that $p>c$. The optimal solution to \eqref{newsvendor} is given by Littlewood's rule $q^*=F^{-1}((p-c)/p)$, where $F^{-1}$ is the quantile function of $D$ (\cite{talluri2006theory}).

\begin{figure}[h]
\centering
\includegraphics[scale = 0.75]{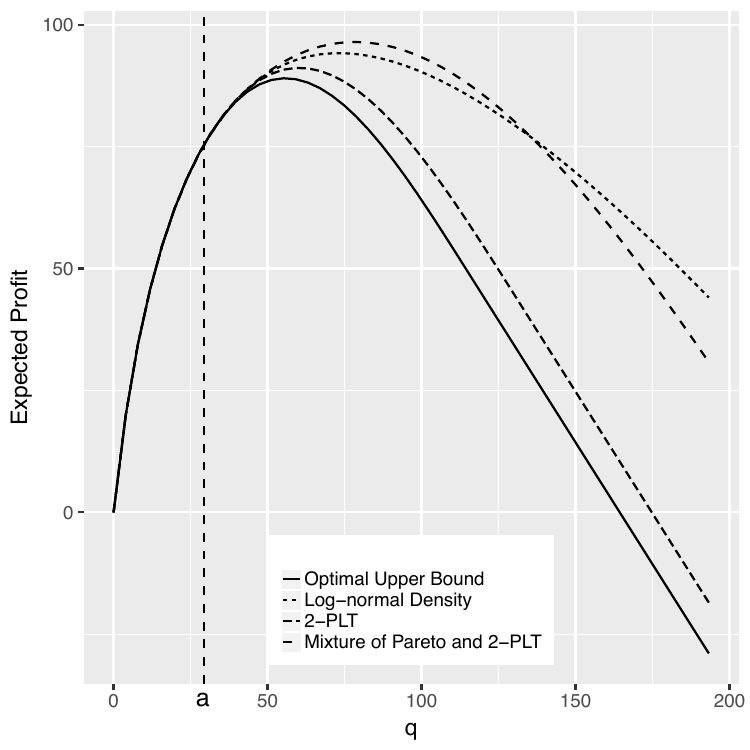}
\caption{\small{Optimal objective values of the inner optimization of the robust newsvendor problem.}}
\label{fig:newsvendor}
\end{figure}

Suppose the distribution of $D$ is only known to have the shape of a lognormal distribution with mean 50 and standard deviation 20 
in the interval $[0,a)$, where $a$ is the $70$-percentile of the lognormal distribution. A robust optimization formulation for \eqref{newsvendor} is
\begin{eqnarray}
&&\max_q\min_{P\in\mathcal A} E[p\min(q,D)]-cq\label{robust newsvendor}\\
&=&\max_q\left\{E[p\min(q,D);D\leq a]+\min_{P\in\mathcal A}E[p\min(q,D);D>a]-cq\right\}\notag
\end{eqnarray}
where $\mathcal A$ denotes the set of convex tails that match the given non-tail region. The outer optimization in \eqref{robust newsvendor} is a concave program. 
We concentrate on the inner optimization. Since $p\min(q,D)$ is a non-decreasing function in $D$ on $[0,\infty)$, its negation is non-increasing, and Assumption \ref{regularity} holds (note that minimization here can be achieved by merely maximizing the negation). Correspondingly, Assumption \ref{H limit} can also be easily checked. We can therefore apply Algorithm 1 (with $\beta = 0.7$, $\eta \approx 0.007 $, and $\nu \approx 0.0003$). 
Figure \ref{fig:newsvendor} shows the optimal lower bound of the inner optimization when $p= 7$, $c= 1$ and $q$ varies between $0$ and $193.26$ (which is the $95$-percentile of the lognormal distribution). The curve peaks at $q = 55.7$, 
which is the solution to problem \eqref{robust newsvendor}.
As a comparison, we also show different candidate values of the expectation that are obtained by fitting the tails of lognormal, 2-PLT (two instances) and mixture of shifted Pareto and 2-PLT (see Figure \ref{fig:newsvendor}). 

\begin{figure}
    \centering
    \begin{subfigure}[b]{0.45\textwidth}
        \includegraphics[scale = 0.6]{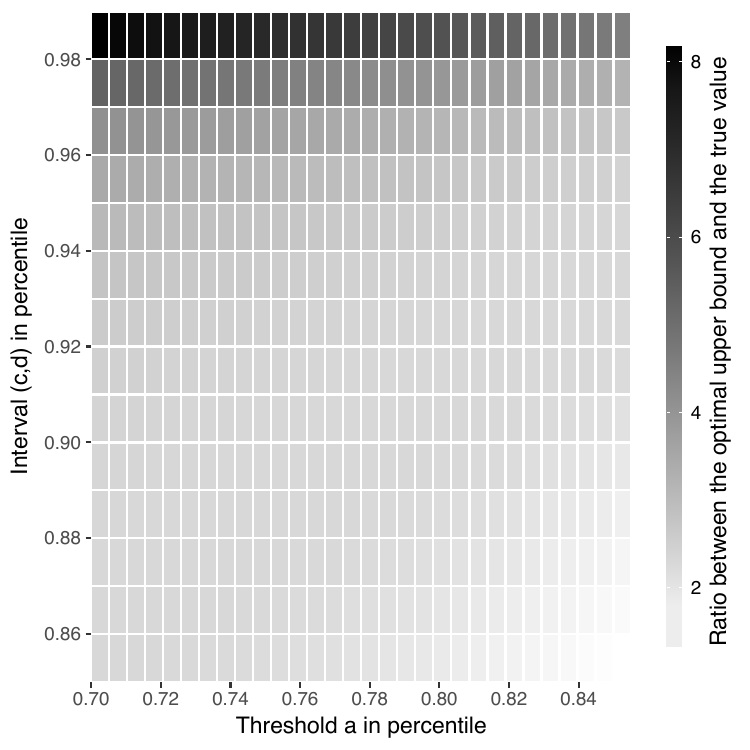}
        \caption{Pareto distribution.}
        \label{fig:tailprob}
    \end{subfigure}
    ~ 
    \begin{subfigure}[b]{0.45\textwidth}
        \includegraphics[scale = 0.6]{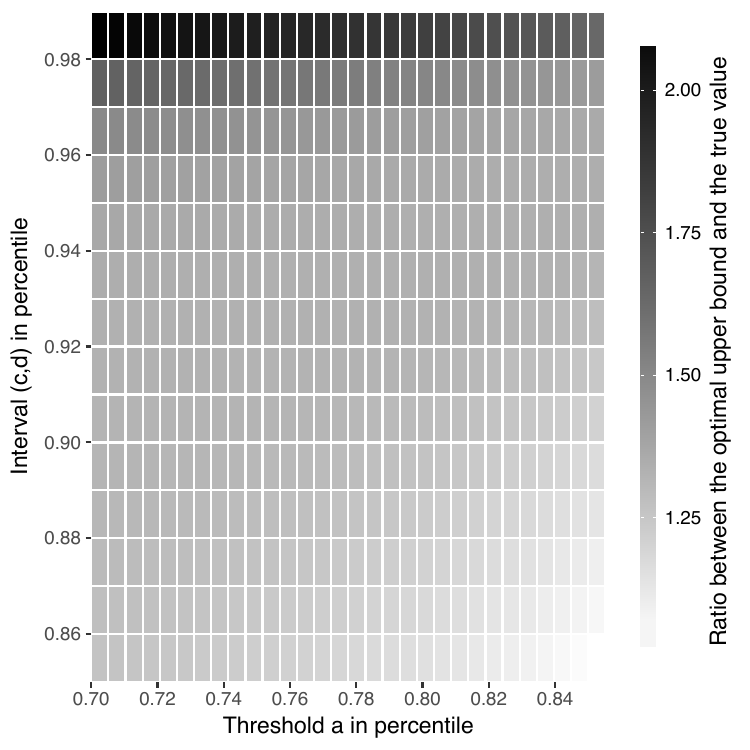}
        \caption{Gamma distribution.}
        \label{fig:tailprobGamma}
    \end{subfigure}
    ~ 
    \caption{Ratio between the worst-case upper bound and the quantity $P(c<X<d)$, at various thresholds $a$ and intervals $(c,d)$ in percentiles, when $X$ follows two different distributions.}
\end{figure}

\noindent\textbf{Tail Interval Probability: }Consider estimating probabilities of the type $P(c<X<d)$. We compare the bound provided by Algorithm 1 with the ``truth" when $X$ is realized from two distributions, a Pareto distribution with tail index 1, i.e., $P(X>x)=1/x$ for all $x>1$, and a Gamma distribution with unit rate and shape parameter $2$, i.e., $P(X>x)=(x+1)e^{-x}$ for all $x>0$. Figures  \ref{fig:tailprob} and \ref{fig:tailprobGamma} give, for various thresholds $a$ in percentile (shown as the $x$-value at the left end of each rectangle) and for various intervals $(c,d)$ also in percentiles (shown as the $y$-values at the lower and upper ends of each rectangle), the ratio between the optimal upper bound and the true probability (represented by the color of each rectangle; the darker the bigger) for these two distributions respectively. We can see that, for the Pareto case, when $a$ is set to the  $70^{th}$ percentile and the interval $(c,d)$ the $(85^{th},86^{th})$-percentiles, the optimal bound given by Algorithm 1 is about twice the truth. For the same threshold $a$ but the interval $(c,d)$ associated with the $(98^{th},99^{th})$-percentiles, the optimal bound is approximately eight times the truth. On the other hand, for the Gamma case, at $a$ equal to the $70^{th}$ percentile and $(c,d)$ the $(98^{th},99^{th})$-percentiles, the bound is at most $2.1$ times the truth. Figures \ref{fig:tailprob} and \ref{fig:tailprobGamma} confirm the intuition that the smaller the distance between $a$ and $c$, the less conservative is the bound. Moreover, the conservativeness level of our generated bound appears to depend on the true distribution. Among the two specifications, our bound is generally tighter when the truth is a Gamma distribution than when it is a Pareto distribution.

\subsection{Synthetic Data: Example \ref{example:synthetic} Revisited}\label{sec:synthetic}
Consider the synthetic data set of size $200$ in Example \ref{example:synthetic}. This data set is actually generated from a lognormal distribution with parameter $(\mu,\sigma)=(0,0.5)$, but we assume that only the data are available to us. We are interested in the quantity $P(4<X<5)$, and for this we will solve program \eqref{opt relaxed} to generate an upper bound that is valid with $95\%$ confidence.

We compute the interval estimates for $\beta$, $\eta$ and $\nu$ as follows. First, we obtain point estimates for these parameters through standard kernel density estimator (KDE) in the R statistical package. To obtain interval estimates, we run $1,000$ bootstrap resamples and take the appropriate quantiles of the $1,000$ resampled point estimates. 
To account for the fact that three parameters are estimated simultaneously, we apply a Bonferroni correction, so that the confidence level used for each individual estimator is $1-0.05/3$.

\begin{figure}[h]
\centering
\includegraphics[scale=.75]{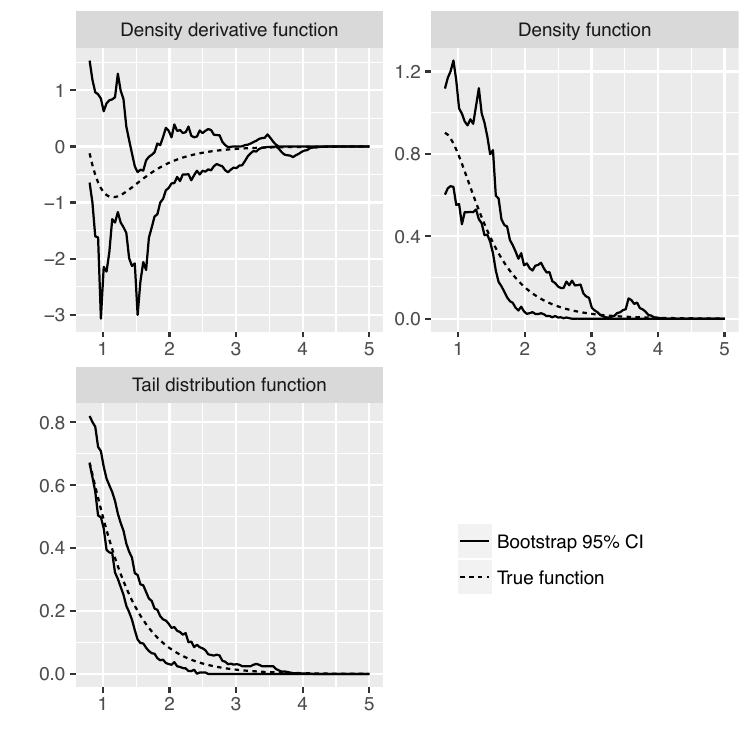}
\caption{Bootstrapped kernel estimation of the distribution, density and density derivative for the synthetic data.\label{Fig:Fit KE}}
\end{figure}

For a sense of how to choose $a$, Figure \ref{Fig:Fit KE} shows the density and density derivative estimates and compares them to those of the lognormal distribution. The KDE suggests that convexity holds starting from around $x=1.5$ (the point where the density derivative estimate starts to turn from a decreasing to an increasing function). Thus, it is reasonable to confine the choice of $a$ to be larger than $1.5$. In fact, this number is quite close to the true inflexion point $1.15$.

Since the data become progressively scanter as $x$ grows larger, and the KDE is designed to utilize neighborhood data, the interval estimators for the necessary parameters $\beta$, $\eta$ and $\nu$ become less reliable for larger choices of $a$. For instance, Figure \ref{Fig:Fit KE} shows that the bootstrapped KDE CI of the density derivative covers the truth only up to $x=3.1$. In general, a good choice of $a$ should be located at a point where there are some data in the neighborhood of $a$, such that the interval estimators for $\beta$, $\eta$ and $\nu$ are reliable, but as large as possible, because choosing a small $a$ can make the tail extrapolation bound more conservative. 

As a first attempt, we run Algorithm 2 using $a=3.1$ to estimate an upper bound for the probability $P(4<X<5)$, which gives $8.8\times10^{-3}$ while the truth is $2.1\times10^{-3}$. Thus, this estimated upper bound does cover the truth and also has the same order of magnitude. We perform the following two other procedures for comparison:
\begin{enumerate}
\item GPD approach: As discussed in Section \ref{Sec:Existing Techniques}, this is a common approach for tail modeling. Fit the data above a threshold $u$ to the density function
$$(1 - \hat F(u))g_{\hat \zeta,\hat \beta}(x-u)$$
where $\hat F(u)$ is the estimated ECDF at $u$, and $g_{\hat \zeta,\hat \beta}(\cdot)$ is the GPD density, whose distribution function is defined as
\begin{equation*}
G_{\zeta,\beta}(x) = \begin{cases}
                        1 - (1+\zeta x/\beta)^{-1/\zeta} & \mbox{if}\quad \zeta \neq 0\\
                        1 - \exp(-x/\beta) & \mbox{if}\quad \zeta = 0
                        \end{cases}
\end{equation*}
for $x\geq0$ if $\zeta\geq0$ and $0\leq x\leq-\beta/\zeta$ if $\zeta<0$, and $\beta>0$. Set the threshold $u$ to be $1.8$, the point at which a linear trend begins to be observed on the mean excess plot of the data, as recommended by \cite{McNeil1997}. Estimate $\hat F(u)$ by the sample mean of $I(X_i\leq u)$, where $I(\cdot)$ denotes the indicator function. Obtain the parameter estimates $\hat\zeta$ and $\hat\beta$ using the maximum likelihood estimator suggested by \cite{smith1987estimating}.
Then use the delta method to obtain a $95\%$ CI of the quantity $P(c<X<d)$.


\item Worst-case approach with known parameter values: Assume $\beta$, $\eta$ and $\nu$ are known at $a=3.1$. Then run Algorithm 1 to obtain the upper bound. 
\end{enumerate}

Table \ref{Tab:Ex SyntD} shows the upper bounds obtained from the above approaches, and also shows the obvious fact that using ECDF alone for estimating $P(4<X<5)$ gives $0$ since there are no data in the interval $[4,5]$. The $95\%$ CI output by GPD fit is $[-8.72\times10^{-4},1.10\times10^{-3}]$, which does not bound the truth (note that this is a two-sided interval, and the upper bound would be off even more if it had been one-sided). 
The worst-case approach with known parameters gives an upper bound of $3.16\times10^{-3}$, which is less conservative than the case when the parameters are estimated. The difference between these numbers can be interpreted as the price of estimation for $\beta$, $\eta$ and $\nu$. For this particular setup, the worst-case approach correctly covers the true value, whereas GPD fitting gives an invalid upper bound, thus showing that either the data size or the threshold level is insufficient to support a good fit of the GPD. This is an instance where the worst-case approach has outperformed GPD in terms of correctness.
\begin{table}[ht]
\centering
\begin{tabular}{cc}
  \hline
Method & Estimated upper bound \\ 
  \hline
Truth & 2.14E-03 \\ 
  ECDF & 0.00E+00 \\ 
  GPD & 1.11E-03 \\ 
  Worst-case with known parameters & 3.16E-03 \\ 
  Worst-case appoach & 8.80E-03 \\ 
   \hline
\end{tabular}
\caption{Estimated upper bounds of the probability $P(4<X<5)$ for the synthetic data in Example \ref{example:synthetic}.} 
\label{Tab:Ex SyntD}
\end{table}

Given that the worst-case approach with estimated parameters appears conceivably more conservative than with known parameters, we conduct a sensitivity study using only Algorithm 1. The first row in Table \ref{Tab:Ex SyntD  sensitivity} shows the upper bound output by Algorithm 1 using the point estimates of the parameters $\beta,\eta,\nu$. The other rows in Table \ref{Tab:Ex SyntD  sensitivity} show the outputs of Algorithm 1 when some values of the parameters are changed to the upper estimates of the $95\%$ CIs. Some scenarios are omitted in the table because they lead to infeasibility. We see that among all these scenarios, the most conservative upper bound occurs when $\beta,\eta,\nu$ are all set to be the upper estimates, giving to $8.67\times10^{-3}$ which is very close to using Algorithm 2. Note that some of these bounds do not cover the truth, which necessitates the use of the interval approach and Algorithm 2.

\begin{table}[ht]
\centering
\begin{tabular}{cccc}
  \hline
$ \eta$ & $ \beta$ & $ \nu$ & Optimal upper bound \\ 
  \hline
Lower bound & Estimated value & Estimated value & 5.76E-06 \\ 
  Lower bound & Estimated value & Upper bound & 5.76E-06 \\ 
  Lower bound & Upper bound & Estimated value & 5.76E-06 \\ 
  Lower bound & Upper bound & Upper bound & 5.76E-06 \\ 
  Upper bound & Estimated value & Estimated value & 3.61E-04 \\ 
  Upper bound & Estimated value & Upper bound & 1.62E-03 \\ 
  Estimated value & Estimated value & Estimated value & 2.04E-03 \\ 
  Estimated value & Estimated value & Upper bound & 2.05E-03 \\ 
  Estimated value & Upper bound & Upper bound & 5.53E-03 \\ 
  Estimated value & Upper bound & Estimated value & 5.53E-03 \\ 
  Upper bound & Upper bound & Estimated value & 8.30E-03 \\ 
  Upper bound & Upper bound & Upper bound & 8.67E-03 \\ 
   \hline
\end{tabular}
\caption{Sensitivity analysis of the optimal upper bound of $P(4<X<5)$ for the synthetic data in Example \ref{example:synthetic}.} 
\label{Tab:Ex SyntD  sensitivity}
\end{table}


The above discussion focuses only on the realization of one data set, which raises the question of whether it holds more generally. Therefore, we obtain an empirical probability of coverage by repeating the following procedure 100 times:
\begin{enumerate}
\item Generate a lognormal sample of size $200$ with parameters $(\mu,\sigma) = (0,0.5)$;
\item Estimate  $\overline \eta$, $\underline \eta$, $\overline \beta$, $\underline \beta$ and $\overline \nu$ at a chosen point $a$ (see below);
\item Use Algorithm 2 to compute the worst-case upper bound of $P(c<X<d)$.
\end{enumerate}
We then estimate the coverage probability of our worst-case upper bound as the proportion of times that Algorithm 2 yields a bound that dominates the true probability $P(c<X<d)$. We repeat this procedure for different $[c,d]$ varying from $[4,5]$ to $[9,10]$, and for two different values of $a$ given by $3.1$ and $2.8$. Tables \ref{Tab:robustnessOptim} and \ref{Tab:robustnessOptim2} show the true probabilities, the mean upper bounds from the $100$ experiments, and the empirical coverage probabilities.

\begin{table}[ht]
\centering
\begin{tabular}{ccccc}
  \hline
c & d & Truth & Mean upper bound & Coverage probability \\ 
  \hline
  4 &   5 & 2.14E-03 & 1.03E-02 & 0.94 \\ 
    5 &   6 & 4.74E-04 & 6.12E-03 & 0.99 \\ 
    6 &   7 & 1.20E-04 & 4.33E-03 & 1.00 \\ 
    7 &   8 & 3.38E-05 & 3.35E-03 & 1.00 \\ 
    8 &   9 & 1.04E-05 & 2.74E-03 & 1.00 \\ 
    9 &  10 & 3.49E-06 & 2.31E-03 & 1.00 \\ 
   \hline
\end{tabular}
\caption{Mean upper bounds and empirical coverage probabilities using worst-case approach with threshold $a = 3.1$.} 
\label{Tab:robustnessOptim}
\end{table}

\begin{table}[ht]
\centering
\begin{tabular}{rrrrr}
  \hline
c & d & Truth & Mean upper bound & Coverage probability \\ 
  \hline
  4 &   5 & 2.14E-03 & 1.31E-02 & 1.00 \\ 
    5 &   6 & 4.74E-04 & 8.26E-03 & 1.00 \\ 
    6 &   7 & 1.20E-04 & 6.04E-03 & 1.00 \\ 
    7 &   8 & 3.38E-05 & 4.76E-03 & 1.00 \\ 
    8 &   9 & 1.04E-05 & 3.92E-03 & 1.00 \\ 
    9 &  10 & 3.49E-06 & 3.34E-03 & 1.00 \\ 
   \hline
\end{tabular}
\caption{Mean upper bounds and empirical coverage probabilities using worst-case approach with threshold $a = 2.8$.} 
\label{Tab:robustnessOptim2}
\end{table}

The coverage probabilities in Tables \ref{Tab:robustnessOptim} and \ref{Tab:robustnessOptim2} are mostly 1, which suggests that our procedure is conservative. For $a=3.1$ and intervals that are close to $a$, i.e. $[c,d]=[4,5]$ and $[5,6]$, the coverage probability is not 1 but rather is close to the prescribed confidence level of $95\%$. Further investigation reveals that our procedure fails to cover the truth only in the case when the joint CI of the parameters $\eta$, $\beta$ and $\nu$ does not contain the true values, which is consistent with the rationale of our method. Although we have not tried lower values of $a$, it is very likely that in those settings the coverage probabilities will stay mostly 1, and the mean upper bounds will increase since the level of conservativeness increases.

As a comparison, Table \ref{Tab:robustnessGPD} shows the results of GPD fit using the threshold $u= 1.8$. Here, all of the coverage probabilities are far from the prescribed level of $95\%$, which suggests that either GPD is the wrong parametric choice to use since the threshold is not high enough, or that the estimation error of its parameters is too large due to the lack of data. (Again, we have used a two-sided $95\%$ CI for the GPD approach here; if we had used a one-sided upper confidence bound, then the upper bounding value would be even lower and the coverage probability would drop further). However, the mean upper bounds using GPD fit do cover the truth in all cases. Since the coverage probability is well below $95\%$, this suggests that the estimation of GPD parameters is highly sensitive to the realization of data.

\begin{table}[ht]
\centering
\begin{tabular}{ccccc}
  \hline
c & d & Truth & Mean upper bound & Coverage probability \\ 
  \hline
  4 &   5 & 2.14E-03 & 3.87E-03 & 0.62 \\ 
    5 &   6 & 4.74E-04 & 1.27E-03 & 0.53 \\ 
    6 &   7 & 1.20E-04 & 5.48E-04 & 0.51 \\ 
    7 &   8 & 3.38E-05 & 2.79E-04 & 0.43 \\ 
    8 &   9 & 1.04E-05 & 1.62E-04 & 0.40 \\ 
    9 &  10 & 3.49E-06 & 1.03E-04 & 0.37 \\ 
   \hline
\end{tabular}
\caption{Mean upper bounds and empirical coverage probabilities using GPD approach.} 
\label{Tab:robustnessGPD}
\end{table}

In summary, Tables \ref{Tab:robustnessOptim}, \ref{Tab:robustnessOptim2} and \ref{Tab:robustnessGPD} show the pros and cons of our worst-case approach and GPD fitting. GPD is on average closer to the true target quantity, but its confidence upper bound can fall short of the prescribed coverage probability (in fact, only between $37$ to $62\%$ of the time it covers the truth in Table \ref{Tab:robustnessGPD}). On the other hand, our approach gives a reasonably tight upper bound when the interval in consideration (i.e. $[c,d]$) is close to the threshold $a$, and tends to be more conservative far out. This is a drawback, but sensibly so, given that the uncertainty of extrapolation increases as it gets farther away from what is known.

Both our worst-case approach and GPD fitting require choosing a threshold parameter. In GPD fitting, it is important to choose a threshold parameter high enough so that the GPD becomes a valid model. GPD fitting, however, is difficult for a small data set when the lack of data prohibits choosing a high threshold. On the other hand, the threshold in our worst-case approach can be chosen much higher, because our method relies on the data below or close to the threshold, but not those far above it. 

\subsection{Fire Insurance Data: Example \ref{example:fire} Revisited}
Consider the fire insurance data in Example \ref{example:fire}. The quantity of interest is the expected payoff of a high-excess policy with reinsurance, given by $h(x) = (x-50)I(50\leq x<200) + 150 I(x\geq200)$. The data set has only seven observations above $50$.

\begin{figure}[h]
\centering
\includegraphics[scale=.75]{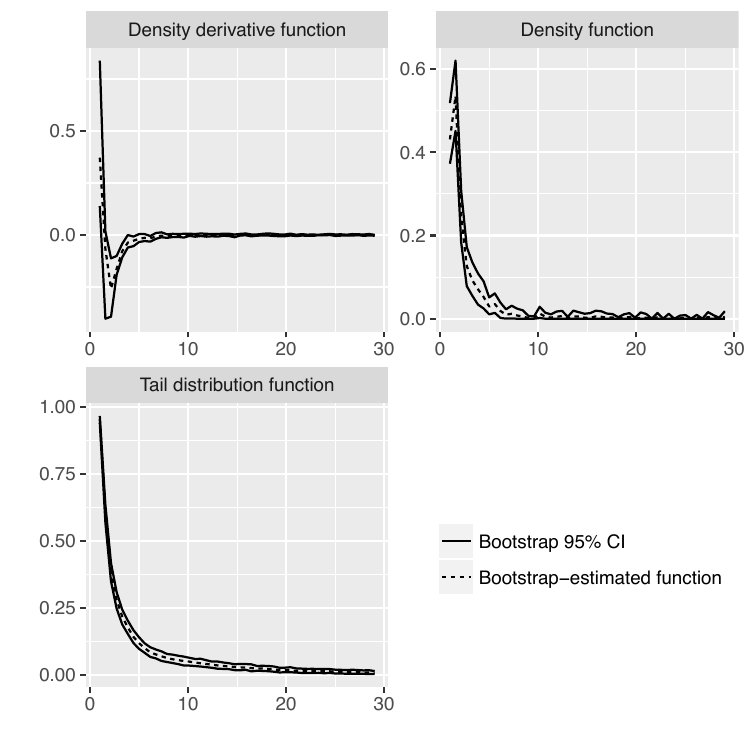}
\caption{Bootstrapped kernel estimation of the distribution, density and density derivative for the the Danish fire losses data in Example \ref{example:fire}.\label{Fig:Fit KE fire}}
\end{figure}

We apply our worst-case approach to estimate an upper bound for the expected payoff by using $a=29.03$, the cutoff above which $15$ observations are available. Similar to Section \ref{sec:synthetic}, we use the bootstrapped KDE to obtain CIs for $\beta$, $\eta$ and $\nu$. The estimates in Figure \ref{Fig:Fit KE fire} appear to be very stable for this example, thanks to the relatively large data size.

We run Algorithm 2 and obtain a $95\%$ confidence upper bound of $1.99$. For comparison, we fit a GPD using threshold $u=10$, which follows \cite{McNeil1997} as the choice that roughly balances the bias-variance tradeoff. The $95\%$ CI from GPD fit is $[-0.03,0.23]$. Thus, the worst-case approach gives an upper bound that is one order of magnitude higher, a finding that resonates with that in Section \ref{sec:synthetic}. Our recommendation is that a modeler who cares only about the order of magnitude would be better off choosing GPD, whereas a more risk-averse modeler who wants a bound on the risk quantity with high probability guarantee would be better off choosing the worst-case approach.

\section{Conclusion}\label{sec:discussions}
This paper proposed a worst-case, nonparametric approach to bound tail quantities based on the tail convexity assumption. The approach relied on an optimization formulated over all possible tail densities. We characterized the optimality structure of this infinite-dimensional optimization problem by developing an equivalence to a moment-constrained problem. Under additional quasi-concavity condition on the objective function, we constructed the numerical solution scheme by converting it into low-dimensional nonlinear programs. With the presence of data, this approach tractably generated statistically valid bounds via suitable relaxations of the optimization that took into account the estimation errors of the required parameters. We compared our proposed approach to existing tail-fitting techniques, and demonstrated its relative strength of outputting correct tail estimates under data-deficient environments. We also examined the level of conservativeness of our bounds, which was viewed as a limitation of the proposed approach.


We suggest two extensions of our research. First is to generalize the proposed method to multivariate distributions, perhaps through separate modeling on the marginal distributions  and the dependency structure. Second is to study means to reduce the level of conservativeness. This can involve mathematical transformations of the variable and the addition of extra information (e.g., other constraints).

\section*{Acknowledgments}
We thank the area editor Bert Zwart, the associate editor and the two referees for many valuable suggestions that have greatly improved the paper. We gratefully acknowledge support from the National Science Foundation under grants CMMI-1400391/1542020 and CMMI-1436247/1523453.

\bibliographystyle{ormsv080}
\bibliography{bibliography1}

\ECSwitch


\ECHead{Appendix}

\section{Proofs for Section \ref{sec:basic}}\label{sec:proofs basic}
We need several results from convex analysis to prove Lemma \ref{thm:reformulation}. For any convex function $g$ on $\mathbb R$, let $\text{dom\ }g=\{x\in\mathbb R:g(x)<\infty\}$ be its effective domain. The following theorems are from \cite{rockafellar2015convex}, specialized to convex functions $g$ with $\text{dom\ }g=\mathbb R$. The definitions of a proper convex function and a closed convex function can be found on p.24 and p.52 therein respectively.

\begin{theorem}[a.k.a. \cite{rockafellar2015convex}, Corollary 10.1.1]
A convex function finite on all of $\mathbb R$ is necessarily continuous.\label{thm1}
\end{theorem}

\begin{theorem}[a.k.a. \cite{rockafellar2015convex}, Theorem 24.1]
Let $g$ be a closed proper convex function on $\mathbb R$, such that $\text{dom}\ g=\mathbb R$. Then $g_+'$ exists and is a finite non-decreasing function on $\mathbb R$. Moreover, $g_+'$ is right-continuous, i.e., $\lim_{z\searrow x}g_+'(z)=g_+'(x)$ for any $x\in\mathbb R$.\label{thm2}
\end{theorem}

\begin{theorem}[a.k.a. \cite{rockafellar2015convex}, Corollary 24.2.1]
Let $g$ be a finite convex function on a non-empty open real interval $I$. Then
$$g(y)-g(x)=\int_x^yg_+'(t)dt$$
for any $x$ and $y$ in $I$.\label{thm3}
\end{theorem}

\begin{theorem}[a.k.a. \cite{rockafellar2015convex}, Theorem 24.2]
Let $\varphi$ be a non-decreasing function from $\mathbb R$ to $[-\infty,\infty]$ such that $\varphi(b)$ is finite for some $b\in\mathbb R$. Then the function given by
$$g(x)=\int_b^x\varphi(t)dt$$
is a well-defined closed proper convex function on $\mathbb R$.\label{thm4}
\end{theorem}

\proof{Proof of Lemma \ref{thm:reformulation}.}
Throughout this proof, without loss of generality let $a=0$ (by replacing $f(x)$ with $f(x+a)$, and $h(x)$ with $h(x+a)$ respectively). Note that optimizations \eqref{opt} and \eqref{opt reformulated} do not depend on $f(x)$ for $x<0$. For the purpose of applying Theorems \ref{thm1}--\ref{thm4} more directly, let us extend $f$ to $\mathbb R^-$, by defining $f(x)=\eta-\nu x$ for $x<0$ (this extension of $f$ is a mathematical artifact and does not necessarily match the given true density).

Let $\mathcal F_1$ be the feasible region in \eqref{opt}, and $\mathcal F_2$ be the feasible region in \eqref{opt reformulated}. We show that $\mathcal F_1=\mathcal F_2$.

\noindent\underline{Proof of $\mathcal F_1\subset\mathcal F_2$: }Since $f(x)<\infty$ for at least one $x\in\mathbb R$ (e.g., take $x=0$) and $f(x)\geq0>-\infty$ for all $x\in\mathbb R$, we get that $f$ is proper (\cite{rockafellar2015convex}, p.24).

Next, we argue that $f(x)<\infty$ for all $x\in\mathbb R$. Suppose on the contrary that $f(x_0)=\infty$ for some $x_0>0$. If $f(y)<\infty$ for some $y>x_0$, then $((y-x_0)/y)f(0)+(x_0/y)f(y)=((y-x_0)/y)\eta+(x_0/y)f(y)<\infty=f(x_0)$, contradicting \eqref{c4}. But if $f(y)=\infty$ for all $y>x_0$, then $\int_0^\infty f(t)dt=\infty$, contradicting \eqref{c1}. Therefore, $f(x)<\infty$ for all $x\in\mathbb R$, and with \eqref{c5}, we conclude that $f$ is finite.

Since $f$ is finite on $\mathbb R$, Theorem \ref{thm1} implies that $f$ is continuous and hence lower semi-continuous. Since $f$ is proper, lower semi-continuity is the same as closedness (\cite{rockafellar2015convex}, p.52). Hence $f$ is closed.

Therefore, together with the convexity condition in \eqref{c4}, Theorem \ref{thm2} implies the existence of $f_+'$ that satisfies \eqref{cc4}. Moreover, Theorem \ref{thm3} implies \eqref{cc6}.

Next, with the monotonicity of $f_+'$ by \eqref{cc4}, we have $f_+'(x)\geq f_+'(0)=-\nu$ for all $x\geq0$, thus implying the first inequality of \eqref{cc3}. To prove the second inequality in \eqref{cc3}, suppose in the contrary that $f_+'(x_0)>0$ for some $x_0>0$. Since $f_+'(x)\geq f_+'(x_0)>0$ for all $x>x_0$ by \eqref{cc4}, we have, from \eqref{cc6}, $f(x)=\int_0^xf_+'(t)dt+\eta\to\infty$, implying that $\int_0^\infty f(x)dx=\infty$ and contradicting \eqref{c1}. Hence the second inequality in \eqref{cc3} holds. We have therefore shown \eqref{cc3}.

Lastly, suppose that $f_+'(x)\not\to0$. Then, since \eqref{cc3} holds, there exists a sequence $x_k\to\infty$ such that $f_+'(x_k)\to c<0$. But since $f_+'$ is monotone by \eqref{cc4}, $\lim_{x\to\infty}f_+'(x)$ exists and must equal $c$. But then, from \eqref{cc6}, $f(x)=\int_0^xf_+'(t)dt+\eta\to-\infty$, violating \eqref{c5}. Thus \eqref{cc5} holds.

The constraints \eqref{cc1} and \eqref{cc2} follow immediately from \eqref{c1} and \eqref{c2}. We therefore conclude that $\mathcal F_1\subset\mathcal F_2$.

\noindent\underline{Proof of $\mathcal F_2\subset\mathcal F_1$: }Since $f_+'$ is bounded on $\mathbb R$ by \eqref{cc3}, Theorem \ref{thm4} and \eqref{cc4} (with $f_+'(x)$ defined as $-\nu$ for $x<0$) implies that the $f$ defined by \eqref{cc6} is convex on $\mathbb R$, giving \eqref{c4}.

Suppose $f(x_0)<0$ for some $x_0>0$. Then, since $f_+'\leq0$ by \eqref{cc3}, \eqref{cc6} implies $f(x)<0$ for all $x\geq x_0$. Thus $\int_0^\infty f(x)dx=-\infty$, contradicting \eqref{cc1}. Therefore, \eqref{c5} holds.

The constraint \eqref{cc3} implies \eqref{c3} immediately. The condition \eqref{cc6} implies $f(0)=f(0+)$. Thus, combining with \eqref{cc2}, we get that \eqref{c2} holds. Finally, note that \eqref{cc1} is the same as \eqref{c1}. We conclude that $\mathcal F_2\subset\mathcal F_1$.\hfill\Halmos
\endproof

To prove Theorem \ref{main reduction}, we first introduce the following lemma:
\begin{lemma}
If $f$ is a feasible solution of \eqref{opt}, equivalently \eqref{opt reformulated}, then $xf(x)\to0$ and $x^2f_+'(x)\to0$ as $x\to\infty$.\label{tail}
\end{lemma}

\proof{Proof of Lemma \ref{tail}.}
We need the observations that $f(x)$ is non-increasing by \eqref{cc3}, $f(x)\geq0$ for all $x\geq a$ by \eqref{c5}, and that $f$ is integrable on $[a,\infty)$ with $\int_a^\infty f(x)dx=\beta$ by \eqref{c1}. Denote $F(x)=\int_a^xf(t)dt$ and $g(x)=xf(x)-F(x)$. Consider, for $a\vee0\leq x_1\leq x_2$,
\begin{align*}
g(x_2) -g(x_1)  &= x_2f(x_2)-x_1f(x_1) - (F(x_2)-F(x_1)) \\
                &\leq x_2f(x_2)-x_1f(x_1) - f(x_2)( x_2  -  x_1) \text{\ \ \ \ since $f(x)$ is non-increasing}\\
                &= x_1[f(x_2) - f(x_1)]\\
                &\leq 0 \text{\ \ \ \ again since $f$ is non-increasing}
\end{align*}
Therefore $g$ is non-increasing for $x\geq a\vee0$, and since $xf(x)\geq0$ and $0\leq F(x)\leq\beta$ for $x\geq a\vee0$, we have $g$ bounded from below on the same range. This implies that $g$ must converge to a limit, say $c$, as $x\to\infty$. In other words, $xf(x)-F(x)\to c$, and since $F(x)\to\beta$, we have $xf(x)\to c+\beta$. Since $xf(x)\geq0$ for $x\geq a\vee0$, there are two cases: $c+\beta>0$ or $c+\beta=0$. The first case implies that $xf(x)\geq\epsilon>0$ for some $\epsilon$ for all large enough $x$. This means $f(x)\geq\epsilon/x$ for all large enough $x$, and hence $\int_a^\infty f(x)dx=\infty$, which contradicts \eqref{c1}. Therefore $xf(x)$ must converge to 0. This proves the first part of the lemma.

To prove the second part, we need the observation that $f_+'(x)$ is non-decreasing for $x\geq a$ by \eqref{cc4}, and is non-positive for $x\geq a$ by \eqref{cc3}. Also, by \eqref{cc6} we have $f(x)=\int_a^xf_+'(t)dt+\eta$ for $x\geq a$. Let $\bar F(x)=\int_x^\infty f(t)dt$ for $x\geq a$, which is finite and converges to 0 by \eqref{c1}. We now define $\tilde g(x) = -x^2f_+'(x) + 2\tilde  F(x)$, where $\tilde F(x) = -\int_x^\infty t f_+'(t)dt$, for $x\geq a$. Note that $xf_+'(x)$ is integrable on $[a,\infty)$ because the absolute continuity of $f$, and $\lim_{x\to\infty}xf(x)\to 0$ as we have just proved, which allows integration by parts yielding
\begin{equation}
\tilde F(x)=-\int_x^\infty t f_+'(t)dt=-tf(t)|_x^\infty+\int_x^\infty f(t)dt=xf(x)+\bar F(x)<\infty\label{interim2new}
\end{equation}

For any $(a\vee0) \leq x_1 \leq x_2$,
\begin{align*}
\tilde g(x_2) - \tilde g(x_1) &= x_1^2f_+'(x_1)- x_2^2f_+'(x_2) - 2\tilde F(x_1) + 2\tilde F(x_2) \\
                &\leq x_1^2f_+'(x_1)- x_2^2f_+'(x_2) + f_+'(x_2) (x_2^2 - x_1^2) \text{\ \ \ \ since $f_+'(x)$ is non-decreasing}\\
                &= x_1^2(f_+'(x_1)-f_+'(x_2))\\
                &\leq0 \text{\ \ \ \ again since $f_+'(x)$ is non-decreasing}
\end{align*}
Therefore, $\tilde g(x)$ is non-increasing for $x\geq a$. Note that $-x^2f_+'(x)\geq0$ for $x\geq a$. Also, from \eqref{interim2new}, since $\lim_{x\to\infty}xf(x)\to 0$ and $\bar F(x)\to0$, we have $\tilde F(x)\to0$ as $x\to\infty$ and hence also bounded for large enough $x$. Therefore $\tilde g$ is bounded from below. This implies that $\tilde g$ must converge to a limit, say $\tilde c$, as $x\to\infty$. Since $\tilde F(x)\to0$, we have $-x^2f_+'(x)\to\tilde c$. Since $-x^2f_+'(x)\geq0$ for $x\geq a$, there are two cases: either $\tilde c>0$ or $\tilde c=0$. The former case implies that $-xf_+'(x)\geq\epsilon/x$ for some $\epsilon>0$ and large enough $x$, and so $\tilde F(x)= -\int_x^\infty xf_+'(x)dx = \infty$ for $x\geq a$, which contradicts \eqref{interim2new}. Therefore $-x^2f_+'(x)\to0$. This proves the second part of the lemma.\hfill\Halmos
\endproof

\proof{Proof of Theorem \ref{main reduction}.}
Throughout this proof, without loss of generality let $a=0$. By Lemma \ref{tail}, we can introduce the extra conditions $xf(x)\to0$ and $x^2f_+'(x)\to0$ as $x\to\infty$ into formulation \eqref{opt reformulated}. In other words, formulation \eqref{opt reformulated} is equivalent to (letting $a=0$)
\begin{subequations}\label{opt reformulated new}
\begin{eqnarray}
&\underset{f}{\max}&\int_0^\infty h(x)f(x)dx\notag\\
&\text{subject to\ \ }&\int_0^\infty f(x)dx=\beta\label{cc1 new}\\
&&f(0)=\eta\label{cc2 new}\\
&&f_+'(x)\text{\ exists and is non-decreasing and right-continuous for\ }x\geq 0\label{cc4 new}\\
&&-\nu\leq f_+'(x)\leq0\text{\ for\ }x\geq 0\label{cc3 new}\\
&&f_+'(x)\to0\text{\ as\ }x\to\infty\label{cc5 new}\\
&&f(x)=\int_0^xf_+'(t)dt+\eta\text{\ for\ }x\geq 0\label{cc6 new}\\
&&xf(x)\to0\text{\ and\ }x^2f_+'(x)\to0\text{\ as\ }x\to\infty\label{cc7 new}
\end{eqnarray}
\end{subequations}

For convenience, we let $\tilde H(x)=\int_0^xh(u)du$ and $H(x)=\int_0^x\tilde H(u)du$. Consider the objective function of \eqref{opt reformulated new}. Since $\tilde H$ is continuous and $f$ is absolutely continuous with $f(x)=\int_0^xf_+'(t)dt+\eta$ by \eqref{cc6 new}, we have, using integration by parts,
\begin{equation}
\int_0^\infty h(x)f(x)dx=\tilde H(x)f(x)\Big|_0^\infty-\int_0^\infty\tilde H(x)f_+'(x)dx=-\int_0^\infty\tilde H(x)f_+'(x)dx\label{interim1new}
\end{equation}
where the second equality follows from \eqref{cc7 new} and that $\tilde H(x)=O(x)$ as $x\to\infty$ since $h$ is bounded. As $H$ is continuous and $f_+'$ has bounded variation by \eqref{cc3 new} and \eqref{cc4 new}, we have, using integration by parts again, that \eqref{interim1new} is equal to
\begin{equation}
-H(x)f_+'(x)\Big|_0^\infty+\int_0^\infty H(x)df_+'(x)=\int_0^\infty H(x)df_+'(x)\label{interim1}
\end{equation}
where the equality follows from \eqref{cc7 new} and that $H(x)=O(x^2)$ as $x\to\infty$ since $h$ is bounded.


For \eqref{cc1 new}, we can write
\begin{equation}
\int_0^\infty f(x)dx=\int_0^\infty\frac{x^2}{2}df_+'(x)\label{interim2}
\end{equation}
by merely viewing $h\equiv1$ in \eqref{interim1new} and \eqref{interim1}. Also, since $f(x)\to0$ as $x\to\infty$ by \eqref{cc7 new}, we can use integration by parts again to write
\begin{equation}
f(0)=-\int_0^\infty f_+'(x)dx=-xf_+'(x)\Big|_0^\infty+\int_0^\infty xdf_+'(x)=\int_0^\infty xdf_+'(x)\label{interim3}
\end{equation}
where the third equality follows from \eqref{cc7 new} again. Therefore, \eqref{opt reformulated new} can be written as
\begin{subequations}\label{opt3}
\begin{eqnarray}
&\underset{f}{\max}&\int_0^\infty H(x)df_+'(x)\notag\\
&\text{subject to\ \ }&\int_0^\infty\frac{x^2}{2}df_+'(x)=\beta\label{cc1 new1}\\
&&\int_0^\infty xdf_+'(x)=\eta\label{cc2 new1}\\
&&f_+'(x)\text{\ exists and is non-decreasing and right-continuous for\ }x\geq0\label{cc4 new1}\\
&&-\nu\leq f_+'(x)\leq0\text{\ for\ }x\geq0\label{cc3 new1}\\
&&f_+'(x)\to0\text{\ as\ }x\to\infty\label{cc5 new1}\\
&&f(x)=\int_0^xf_+'(t)dt+\eta\text{\ for\ }x\geq 0\label{cc6 new1}\\
&&xf(x)\to0\text{\ and\ }x^2f_+'(x)\to0\text{\ as\ }x\to\infty\label{cc7 new1}
\end{eqnarray}
\end{subequations}
We show that \eqref{cc7 new1} is redundant. By \eqref{cc1 new1}, we have $\int_0^\infty(x^2/2)df_+'(x)<\infty$ and hence $\int_x^\infty(t^2/2)df_+'(t)\to0$ as $x\to\infty$. Now, for $x\geq0$, we have
$$\int_x^\infty\frac{t^2}{2}df_+'(t)\geq\frac{x^2}{2}\int_x^\infty df_+'(t)=-\frac{x^2}{2}f_+'(x)\geq0$$
where the first inequality follows since $f_+'(x)$ is non-decreasing by \eqref{cc4 new1}, the equality follows from \eqref{cc5 new1}, and the last inequality from \eqref{cc3 new1}. Therefore, $-(x^2/2)f_+'(x)\to0$ as $x\to\infty$. This shows that the second part of \eqref{cc7 new1} is redundant.

By \eqref{cc2 new1}, and since $f_+'(x)$ is monotone, we can use integration by parts to get
\begin{equation}
\eta=\int_0^\infty xdf_+'(x)=xf_+'(x)\Big|_0^\infty-\int_0^\infty f_+'(x)dx=-\int_0^\infty f_+'(x)dx\label{new equation1}
\end{equation}
where the last equality follows since we have proved $-(x^2/2)f_+'(x)\to0$ and so $xf_+'(x)\to0$ as $x\to\infty$. Now, using \eqref{cc6 new1} and \eqref{new equation1}, we write
\begin{equation}
f(x)=\int_0^xf_+'(t)dt+\eta=\int_0^xf_+'(t)dt-\int_0^\infty f_+'(t)dt=-\int_x^\infty f_+'(t)dt\label{new equation new1}
\end{equation}
Since $-(x^2/2)f_+'(x)\to0$ as $x\to\infty$, we have $f_+'(x)=o(1/x^2)$. So $-\int_x^\infty f_+'(t)dt=o(1/x)$. Then \eqref{new equation new1} implies the first part of \eqref{cc7 new1} is redundant.

Therefore, \eqref{opt3} can be written as
\begin{equation}
\begin{array}{ll}
\underset{f}{\max}&\int_0^\infty H(x)df_+'(x)\\
\text{subject to\ \ }&\int_0^\infty\frac{x^2}{2}df_+'(x)=\beta\\
&\int_0^\infty xdf_+'(x)=\eta\\
&f_+'(x)\text{\ exists and is non-decreasing and right-continuous for\ }x\geq0\\
&-\nu\leq f_+'(x)\leq0\text{\ for\ }x\geq0\\
&f_+'(x)\to0\text{\ as\ }x\to\infty
\end{array}\label{opt3 new}
\end{equation}
and the constraint \eqref{cc6 new1} in \eqref{opt reformulated new} states that $f$ can be recovered from $f(x)=\int_0^xf_+'(t)dt+\eta$. Note that this definition of $f$ must necessarily have a right derivative coinciding with the obtained $f_+'(x)$.

Finally, let $p(x)=f_+'(x)/\nu+1$. Then \eqref{opt3 new} can be rewritten as
\begin{equation}
\begin{array}{ll}
\underset{p}{\max}&\nu\int_0^\infty H(x)dp(x)\\
\text{subject to\ \ }&\int_0^\infty x^2dp(x)=\frac{2\beta}{\nu}\\
&\int_0^\infty xdp(x)=\frac{\eta}{\nu}\\
&p(x)\text{\ non-decreasing and right-continuous for\ }x\geq 0\\
&0\leq p(x)\leq1\text{\ for\ }x\geq0\\
&p(x)\to1\text{\ as\ }x\to\infty
\end{array}\label{opt4}
\end{equation}
or equivalently
\begin{equation}
\begin{array}{ll}
\underset{p}{\max}&\nu\int_{-\infty}^\infty H(x)dp(x)\\
\text{subject to\ \ }&\int_{-\infty}^\infty x^2dp(x)=\frac{2\beta}{\nu}\\
&\int_{-\infty}^\infty xdp(x)=\frac{\eta}{\nu}\\
&p(x)\text{\ non-decreasing and right-continuous for\ }x\in\mathbb R\\
&0\leq p(x)\leq1\text{\ for\ }x\in\mathbb R\\
&p(x)\to1\text{\ as\ }x\to\infty\\
&p(x)=0\text{\ for\ }x<0
\end{array}\label{opt new}
\end{equation}
since $H(x)=x=x^2=0$ at $x=0$. One can uniquely identify, up to measure zero, a non-decreasing, right-continuous $p$ such that $\lim_{x\to\infty}p(x)=1$ and $p(x)=0$ for $x<0$ with a probability measure supported on $\mathbb R^+$. Hence \eqref{opt new} is equivalent to \eqref{prob opt main}. This concludes the result.\hfill\Halmos
\endproof

To prove Theorem \ref{equivalence finite}, we need several results from \cite{winkler1988extreme} stated below.

\begin{theorem}[\cite{winkler1988extreme} Theorem 2.1(b)]
Let $\mathcal X$ be a measurable space with $\sigma$-field $\mathcal F$ and suppose that $\mathcal P$ is a simplex of probability measures whose extreme points are Dirac measures. Fix measurable functions $f_1,\ldots,f_n$ and real numbers $c_1,\ldots,c_n$. Consider the set
$$\mathcal H=\left\{q\in\mathcal P:f_i\text{\ is $q$-integrable and\ }\int f_idq=c_i,\ 1\leq i\leq n\right\}$$
Then $\mathcal H$ is convex and
\begin{eqnarray*}
\text{ex\ }\mathcal H&=\Bigg\{&q\in\mathcal H:q=\sum_{i=1}^mt_i\cdot\delta(x_i),\ t_i>0,\ \sum_{i=1}^mt_i=1,\ x_i\in\mathcal X,\ 1\leq m\leq n+1,\\
&&\text{the vectors\ }(f_1(x_i),\ldots,f_n(x_i),1),\ 1\leq i\leq m,\text{\ are linearly independent}\Bigg\}
\end{eqnarray*}
where $\text{ex\ }\mathcal H$ denotes the set of all extreme points of $\mathcal H$.\label{thm11}
\end{theorem}

\begin{theorem}[Adapted from \cite{winkler1988extreme} Theorem 3.2]
Let $\mathcal X$ be a Hausdorff space, $\mathcal F$ be the Borel $\sigma$-field and $\mathcal P_r(\mathcal X)$ be the set of regular probability measures on $\mathcal X$. Let
$$\mathcal H=\left\{q\in\mathcal P_r(\mathcal X):f_i\text{\ is $q$-integrable and\ }\int f_idq=c_i,\ 1\leq i\leq n\right\}$$
Every measure affine functional $J$ on $\mathcal H$ fulfills
$$\sup\{J(q):q\in\mathcal H\}=\sup\{J(q):q\in\text{ex\ }\mathcal H\}$$\label{thm21}
\end{theorem}

Theorem \ref{thm21} is precisely Theorem 3.2 in \cite{winkler1988extreme}, except replacing the inequalities with equalities for the moments that define $\mathcal H$, which is immediate (and is pointed out by the comment right after the theorem in \cite{winkler1988extreme}).

\begin{proposition}[\cite{winkler1988extreme} Proposition 3.1]
Let $\mathcal X$, $\mathcal F$ and $\mathcal H$ be given as in Theorem \ref{thm21} and the function $g$ on $\mathcal X$ be integrable for every $q\in\mathcal H$ (possibly with integral values $\infty$ or $-\infty$). Then the functional $G$ on $\mathcal H$ defined by $G(q)=\int_{\mathcal X}gdq$ is measure affine.\label{thm31}
\end{proposition}

\proof{Proof of Theorem \ref{equivalence finite}.}
By Examples 2.1(a) in \cite{winkler1988extreme}, the set $\mathcal P$ in Theorem \ref{thm11} can be chosen to be the set of all regular probability measures. On Polish space every probability measure is regular. Therefore, on the space $\mathbb R^+$, which is Polish, we can take $\mathcal P$ in Theorem \ref{thm11} as the set of all probability measures. The $\mathcal H$ in Theorems \ref{thm11} and \ref{thm21} then coincide. By Proposition \ref{thm31}, the objective $\nu\mathbb E[H(X)]$ in $OPT(\mathcal P^+)$ is measure affine. Therefore, using Theorems \ref{thm11} and \ref{thm21}, and noting that
$$\mathcal H\supseteq\left\{q\in\mathcal H:q=\sum_{i=1}^mt_i\cdot\delta(x_i),\ t_i>0,\ \sum_{i=1}^mt_i=1,\ x_i\in\mathcal X,\ 1\leq m\leq n+1\right\}\supseteq\text{ex\ }\mathcal H$$
for the coincided $\mathcal H$ in Theorems \ref{thm11} and \ref{thm21}, we conclude the theorem.\hfill\Halmos
\endproof


\proof{Proof of Proposition \ref{char}.}
If program \eqref{prob opt main} is consistent, then by Theorem \ref{equivalence finite}, either an optimal solution in $\mathcal P_3^+$ exists, which corresponds to the first case of the lemma, or there exists a feasible sequence $\mathbb P^{(k)}\in\mathcal P_3^+$ such that $Z(\mathbb P^{(k)})\to Z^*$. Let $\mathbb P^{(k)}\sim(x_1^{(k)},x_2^{(k)},x_3^{(k)},p_1^{(k)},p_2^{(k)},p_3^{(k)})$. Suppose that $x_i$'s are all bounded above by a number, say $M$. Then, since $[0,M]^3\times\mathcal S_3$ is a compact set, by Bolzano-Weierstrass Theorem we must have a subsequence of $(x_1^{(k)},x_2^{(k)},x_3^{(k)},p_1^{(k)},p_2^{(k)},p_3^{(k)})$, say $(x_1^{(k_j)},x_2^{(k_j)},x_3^{(k_j)},p_1^{(k_j)},p_2^{(k_j)},p_3^{(k_j)})$ converge to $(x_1^*,x_2^*,x_3^*,p_1^*,p_2^*,p_3^*)$ in $[0,M]^3\times\mathcal S_3$. Since $H(x)$ is continuous by construction, we have $Z(\mathbb P^{(k_j)})=\nu\sum_{i=1}^3H(x_i^{(k_j)})p_i^{(k_j)}\to\nu\sum_{i=1}^3H(x_i^*)p_i^*=Z(\mathbb P^*)$, where $\mathbb P^*\sim(x_1^*,x_2^*,x_3^*,p_1^*,p_2^*,p_3^*)$. As $Z(\mathbb P^{(k_j)})$ is a subsequence of $Z(\mathbb P^{(k)})$, $Z(\mathbb P^*)$ must be equal to $Z^*$, and so $\mathbb P^*$ is an optimal solution, which reduces to the first case in the lemma. Therefore, for the second case, we should focus on the scenario that at least one $x_i^{(k)}$ satisfies $\limsup_{k\to\infty}x_i^{(k)}=\infty$.

Without loss of generality, we fix the convention that $x_1^{(k)}\leq x_2^{(k)}\leq x_3^{(k)}$. If at least one of $x_i^{(k)}$ satisfies $\limsup_{k\to\infty}x_i^{(k)}=\infty$, we must have $\limsup_{k\to\infty}x_3^{(k)}=\infty$. In order that $\mathbb P^{(k)}$ is feasible, $\mathbb E^{(k)}[X]=\mu$ holds and so $x_1^{(k)}\leq\mu$ for all $k$. We now distinguish two cases: either $x_2^{(k)}$ is uniformly bounded, say by a large number $M\geq\mu$, or $\limsup_{k\to\infty}x_2^{(k)}=\infty$ also. Consider the first case. First, we find a subsequence $x_3^{(k_j)}\nearrow\infty$. Since $(x_1^{(k_j)},x_2^{(k_j)})\in[0,M]^2$ which is compact, we can choose a further subsequence $k_{j'}$ such that $(x_1^{(k_{j'})},x_2^{(k_{j'})},x_3^{(k_{j'})})\to(x_1^*,x_2^*,\infty)$ where $(x_1^*,x_2^*) \in [0,M]^2$. Now, since $(p_1^{(k_{j'})},p_2^{(k_{j'})},p_3^{(k_{j'})})\in\mathcal S_3$ which is also compact, we can choose another further subsequence $k_{j''}$ such that $(p_1^{(k_{j''})},p_2^{(k_{j''})},p_3^{(k_{j''})})\to(p_1^*,p_2^*,p_3^*)\in \mathcal S_3$. Note that by the constraint $\mathbb E^{(k)}[X^2]=p_1^{(k_{j''})}{x_1^{(k_{j''})}}^2+p_2^{(k_{j''})}{x_2^{(k_{j''})}}^2+p_3^{(k_{j''})}{x_3^{(k_{j''})}}^2=\sigma$, we must have
$p_3^{(k_{j''})}=(\sigma-p_1^{(k_{j''})}{x_1^{(k_{j''})}}^2-p_2^{(k_{j''})}{x_2^{(k_{j''})}}^2)/{x_3^{(k_{j''})}}^2\leq\sigma/{x_3^{(k_{j''})}}^2\to0$. In conclusion, in this case, we end up being able to find a sequence of measures ${\mathbb P^{(k)}}'\sim({x_1^{(k)}}',{x_2^{(k)}}',{x_3^{(k)}}',{p_1^{(k)}}',{p_2^{(k)}}',{p_3^{(k)}}')$ with $({x_1^{(k)}}',{x_2^{(k)}}',{x_3^{(k)}}',{p_1^{(k)}}',{p_2^{(k)}}',{p_3^{(k)}}')\to(x_1^*,x_2^*,\infty,p_1^*,p_2^*,0)$ where $x_1^*,x_2^*\in\mathbb R^+$ and $(p_1^*,p_2^*)\in\mathcal S_2$.

For the second case, namely when $\limsup_{k\to\infty}x_i^{(k)}=\infty$ for both $i=2$ and $3$. We can argue similarly that there is a sequence of measures ${\mathbb P^{(k)}}'\sim({x_1^{(k)}}',{x_2^{(k)}}',{x_3^{(k)}}',{p_1^{(k)}}',{p_2^{(k)}}',{p_3^{(k)}}')$, such that ${x_2^{(k)}}',{x_3^{(k)}}'\to\infty$ and ${p_2^{(k)}}',{p_3^{(k)}}'\to0$. In other words, $({x_1^{(k)}}',{x_2^{(k)}}',{x_3^{(k)}}',{p_1^{(k)}}',{p_2^{(k)}}',{p_3^{(k)}}')\to(x_1^*,\infty,\infty,1,0,0)$ where $x_1^*\in\mathbb R^+$.\hfill\Halmos
\endproof

\proof{Proof of Lemma \ref{consistency}.}
It follows from Jensen's inequality that for any $\mathbb P\in\mathcal P^+$, $\mathbb E[X^2]\geq\mathbb E[X]^2$, which gives $\sigma\geq\mu^2$ in \eqref{prob opt main}. On the other hand, if $\sigma\geq\mu^2$, it is also rudimentary to find $\mathbb P\in\mathcal P_2^+$ with $\mathbb E[X]=\mu$ and $\mathbb E[X^2]=\sigma$. Substituting $\mu=\eta/\nu$ and $\sigma=2\beta/\nu$, we get $\eta^2\leq2\beta\nu$. Lastly, $\mathbb E[X^2]=\mathbb E[X]^2$ if and only if $\mathbb P$ is a point mass. The equivalent statements regarding program \eqref{opt} follows from Theorem \ref{thm:prob opt}.\hfill\Halmos
\endproof

\proof{Proof of Proposition \ref{counter example}.}
Consider a sequence $f^{(k)}(x),x\geq a$ given by
$$f^{(k)}(x)=\left\{\begin{array}{ll}
\eta-\nu(x-a)&\text{\ for\ }a\leq x\leq x_1^{(k)}+a\\
\eta-\nu x_1^{(k)}-\nu p_2^{(k)}(x-a-x_1^{(k)})&\text{\ for\ }x_1^{(k)}+a\leq x\leq x_2^{(k)}+a\\
0&\text{\ for\ }x_2^{(k)}+a\leq x
\end{array}\right.$$
where
\begin{equation}
\begin{array}{l}
x_1^{(k)}=\mu - \gamma^{(k)}\text{\ and\ }\gamma^{(k)}= \frac{\sigma-\mu^2}{ x_2^{(k)} - \mu}\\
x_2^{(k)}\to\infty\\
p_1^{(k)}=1-p_2^{(k)}\\
p_2^{(k)}=\frac{\sigma-\mu^2}{{x_2^{(k)}}^2-2\mu x_2^{(k)}+\sigma}
\end{array}\label{interim2 counter example}
\end{equation}
and $\mu,\sigma$ are defined in \eqref{def new}.

We claim that $f^{(k)}$ is feasible for \eqref{opt} for large enough $k$. This can be argued by verifying that $(x_1^{(k)},x_2^{(k)},p_1^{(k)},p_2^{(k)})\in\mathcal P_2^+$ is feasible for \eqref{prob opt main} and invoking the one-to-one correspondence between the feasible solutions in \eqref{opt} and \eqref{prob opt main} depicted in Theorem \ref{main reduction}. Here we provide an alternate direct verification. It is obvious that for large enough $x_2^{(k)}$, $f^{(k)}$ is non-negative and convex. Moreover, $f^{(k)}(a)=f^{(k)}(a+)=\eta$ and ${f_+^{(k)}}'(a)\geq-\nu$. To show $\int_a^\infty f(x)dx=\beta$, we first verify that
\begin{equation}
p_1^{(k)}x_1^{(k)}+p_2^{(k)}x_2^{(k)}=\mu\label{relation18}
\end{equation}
and
\begin{equation}
p_1^{(k)}{x_1^{(k)}}^2+p_2^{(k)}{x_2^{(k)}}^2=\sigma\label{relation19}
\end{equation}
for large $k$. In fact, we will do so by showing that $\gamma^{(k)}$ and $p_2^{(k)}$ displayed in \eqref{interim2 counter example} are the unique choices that satisfy \eqref{relation18} and \eqref{relation19} and also $x_1^{(k)}=\mu-\gamma^{(k)}$ and $p_1^{(k)}=1-p_2^{(k)}$. With the latter conditions, \eqref{relation18} and \eqref{relation19} can be written as
$$(1-p_2^{(k)})(\mu-\gamma^{(k)})+p_2^{(k)}x_2^{(k)}=\mu$$
and
$$(1-p_2^{(k)})(\mu-\gamma^{(k)})^2+p_2^{(k)}{x_2^{(k)}}^2=\sigma$$
respectively, which further gives
\begin{equation}
p_2^{(k)}\left(\gamma^{(k)} + x_2^{(k)} -\mu\right) -\gamma^{(k)}=0\label{relation20}
\end{equation}
and
\begin{equation}
p_2^{(k)}\left({x_2^{(k)}}^2 -\left(\mu-\gamma^{(k)}\right)^2 \right) + \left(\mu-\gamma^{(k)}\right)^2 =\sigma\label{relation21}
\end{equation}
From \eqref{relation20} we have
\begin{equation}
p_2^{(k)}=\frac{\gamma^{(k)}}{\gamma^{(k)} + x_2^{(k)} -\mu}\label{relation22}
\end{equation}
Putting \eqref{relation22} into \eqref{relation21}, we get
$$\frac{\gamma^{(k)}}{\gamma^{(k)} + x_2^{(k)} -\mu}\left({x_2^{(k)}}^2 -\left(\mu-\gamma^{(k)}\right)^2 \right) + \left(\mu-\gamma^{(k)}\right)^2 =\sigma$$
which can be simplified to
$$\gamma^{(k)}\left(x_2^{(k)} +  \mu-\gamma^{(k)} \right) + \left(\mu-\gamma^{(k)}\right)^2 =\sigma$$
giving
\begin{equation}
\gamma^{(k)}=\frac{\sigma-\mu^2}{x_2^{(k)}-\mu}\label{relation24}
\end{equation}
Plugging \eqref{relation24} into \eqref{relation22}, we have
\begin{equation}
p_2^{(k)}=\frac{\sigma-\mu^2}{(\sigma-\mu^2) + \left(x_2^{(k)}-\mu\right)^2}\label{relation23}
\end{equation}
thus recovering $\gamma^{(k)}$ and $p_2^{(k)}$ in \eqref{interim2 counter example}.

Therefore,
\begin{align*}
\int_a^\infty f^{(k)}(x)dx&=\int_a^{x_1^{(k)}+a}[\eta-\nu(x-a)]dx+\int_{x_1^{(k)}+a}^{x_2^{(k)}+a}[\eta-\nu x_1^{(k)}-\nu p_2^{(k)}(x-a-x_1^{(k)})]dx\\
&=\eta x_2^{(k)}-\frac{\nu}{2}{x_1^{(k)}}^2-\nu x_1^{(k)}(x_2^{(k)}-x_1^{(k)})-\frac{\nu p_2^{(k)}}{2}(x_2^{(k)}-x_1^{(k)})^2\\
&=\eta x_2^{(k)}+\frac{\nu p_1^{(k)}}{2}{x_1^{(k)}}^2+\frac{\nu p_2^{(k)}}{2}{x_2^{(k)}}^2-\nu p_1^{(k)}x_1^{(k)}x_2^{(k)}-\nu p_2^{(k)}{x_2^{(k)}}^2\text{\ using $p_1^{(k)}=1-p_2^{(k)}$}\\
&=\eta x_2^{(k)}+\frac{\nu\sigma}{2}-\nu x_2^{(k)}\mu\text{\ using \eqref{relation18} and \eqref{relation19}}\\
&=\beta\text{\ using $\eta-\nu\mu=0$ and $\beta=\nu\sigma/2$}
\end{align*}
Hence $f^{(k)}$ is feasible for \eqref{opt} for large enough $k$.

Now, the objective value evaluated at $f^{(k)}$ is
\begin{equation}
\int_a^{x_1^{(k)}+a}h(x)(\eta-\nu(x-a))dx+\int_{x_1^{(k)}+a}^{x_2^{(k)}+a}h(x)(\eta-\nu x_1^{(k)}-\nu p_2^{(k)}(x-a-x_1^{(k)}))dx\label{interim counter example}
\end{equation}
The first term in \eqref{interim counter example} is bounded since $x_1^{(k)}\to\mu$. We focus on the second term. By the assumption, we can find $C>0$ such that $h(x)\geq Cx^\epsilon$ for all $x\geq a$. Then, for large enough $k$,
\begin{eqnarray}
&&\int_{x_1^{(k)}+a}^{x_2^{(k)}+a}h(x)(\eta-\nu x_1^{(k)}-\nu p_2^{(k)}(x-a-x_1^{(k)}))dx\notag\\
&\geq&C\int_{x_1^{(k)}+a}^{x_2^{(k)}+a}x^\epsilon(\eta-\nu x_1^{(k)}-\nu p_2^{(k)}(x-a-x_1^{(k)}))dx\notag\\
&\geq&C\int_{x_1^{(k)}+a}^{x_2^{(k)}+a}[(\eta-\nu p_1^{(k)}x_1^{(k)}+\nu p_2^{(k)}a)x^\epsilon-\nu p_2^{(k)}x^{\epsilon+1}]dx\notag\\
&=&(\eta-\nu p_1^{(k)}x_1^{(k)}+\nu p_2^{(k)}a)\frac{x^{\epsilon+1}}{\epsilon+1}\Big|_{x_1^{(k)}+a}^{x_2^{(k)}+a}-\nu p_2^{(k)}\frac{x^{\epsilon+2}}{\epsilon+2}\Big|_{x_1^{(k)}+a}^{x_2^{(k)}+a}\notag\\
&=&(\eta-\nu p_1^{(k)}x_1^{(k)}+\nu p_2^{(k)}a)\frac{(x_2^{(k)}+a)^{\epsilon+1}}{\epsilon+1}-(\eta-\nu p_1^{(k)}x_1^{(k)}+\nu p_2^{(k)}a)\frac{(x_1^{(k)}+a)^{\epsilon+1}}{\epsilon+1}{}\notag\\
&&{}-\nu p_2^{(k)}\frac{(x_2^{(k)}+a)^{\epsilon+2}}{\epsilon+2}+\nu p_2^{(k)}\frac{(x_1^{(k)}+a)^{\epsilon+2}}{\epsilon+2}\label{interim1 counter example}
\end{eqnarray}
Note that since $p_1^{(k)}\to1$, $x_1^{(k)}\to\mu$, $p_2^{(k)}\to0$ and $\eta-\nu\mu=0$, the second term in \eqref{interim1 counter example} converges to 0. Moreover, since $p_2^{(k)}\to0$, the fourth term also converges to 0. Consider the first term in \eqref{interim1 counter example}. In particular,
\begin{align*}
\eta-\nu p_1^{(k)}x_1^{(k)}+\nu p_2^{(k)}a&=\eta-\nu(1-p_2^{(k)})(\mu-\gamma^{(k)})+\nu p_2^{(k)}a\\
&=p_1^{(k)}\nu\gamma^{(k)}+\nu p_2^{(k)}(\mu+a)
\end{align*}
by using $\eta-\nu\mu=0$ and $p_1^{(k)}=1-p_2^{(k)}$. Substituting $\gamma^{(k)}=(\sigma-\mu^2)/(x_2^{(k)} - \mu)$ and $p_2^{(k)}=\Theta(1/{x_2^{(k)}}^2)$, and using $p_1^{(k)}\to1$, we have
$$(\eta-\nu p_1^{(k)}x_1^{(k)}+\nu p_2^{(k)}a)\frac{(x_2^{(k)}+a)^{\epsilon+1}}{\epsilon+1}=(p_1^{(k)}\nu\gamma^{(k)}+\nu p_2^{(k)}(\mu+a))\frac{(x_2^{(k)}+a)^{\epsilon+1}}{\epsilon+1}=\frac{\nu(\sigma-\mu^2){x_2^{(k)}}^\epsilon}{\epsilon+1}(1+o(1))$$
On the other hand, for the third term in \eqref{interim1 counter example}, substituting $p_2^{(k)}=(\sigma-\mu^2)/({x_2^{(k)}}^2-2\mu x_2^{(k)}+\sigma)$, we have
$$-\nu p_2^{(k)}\frac{(x_2^{(k)}+a)^{\epsilon+2}}{\epsilon+2}=-\frac{\nu(\sigma-\mu^2){x_2^{(k)}}^\epsilon}{\epsilon+2}(1+o(1))$$
Thus, \eqref{interim1 counter example} is equal to
$$\left(\frac{1}{\epsilon+1}-\frac{1}{\epsilon+2}\right)\nu(\sigma-\mu^2){x_2^{(k)}}^\epsilon(1+o(1))\to\infty$$
and hence the optimal value of \eqref{opt} is $\infty$.\hfill\Halmos
\endproof

%
%
%

\section{Proofs for Section \ref{sec:opt procedure}}\label{proofs opt procedure}
To prove Proposition \ref{2-supp}, we borrow the following result:
\begin{lemma}[Adapted from Theorem 5.1 in \cite{birge1991bounding}]
Consider $OPT(\mathcal P[0,\tilde c])$ for any $0<\tilde c<\infty$. Suppose $H$ is convex with derivative $H'$ convex on $(0,c)$ and concave on $(c,\tilde c)$ for some $0\leq c\leq\tilde c$. If $OPT(\mathcal P[0,\tilde c])$ is consistent, then an optimal solution exists and lies in $\mathcal P_2[0,\tilde c]$.\label{2-supp prelim}
\end{lemma}
This lemma follows from Theorem 5.1 in \cite{birge1991bounding} that applies to the associated dual problem.

\proof{Proof of Proposition \ref{2-supp}.}
By Theorem \ref{equivalence finite}, $OPT(\mathcal P^+)$ has the same optimal value as $OPT(\mathcal P_3^+)$. By Lemma \ref{2-supp prelim}, for every $\mathbb P$ feasible for $OPT(\mathcal P_3^+)$, which necessarily has bounded support say on $[0,M]$ for some $M>0$, there exists $\mathbb P'\in\mathcal P_2[0,M]$ with the same first and second moments such that $Z(\mathbb P')\geq Z(\mathbb P)$. Since $\mathcal P_2^+\subset\mathcal P_3^+$, this implies that $OPT(\mathcal P_3^+)$ has the same optimal value as $OPT(\mathcal P_2^+)$, which concludes the proposition.\hfill\Halmos
\endproof

\proof{Proof of Proposition \ref{char1}.}
\underline{Proof of \ref{case1}: }Let the optimal probability measure in $\mathcal P_2^+$ be represented by $(x_1,x_2,p_1,p_2)$. Note that $x_1\neq x_2$ since otherwise $\sigma=\mu^2$. Adopting a similar line of analysis as in \cite{birge1991bounding}, we let $x_1<x_2$ without loss of generality. For a two-support-point distribution to be feasible, we must have $x_1<\mu$. Feasibility also enforces that $p_1x_1+p_2x_2=\mu$, $p_1x_1^2+p_2x_2^2=\sigma$ and $p_1+p_2=1$. Hence $p_2=1-p_1$, which gives $p_1x_1+(1-p_1)x_2=\mu$ and $p_1x_1^2+(1-p_1)x_2^2=\sigma$. From the first equation we get $p_1=(x_2-\mu)/(x_2-x_1)$. Putting this into $p_1x_1^2+(1-p_1)x_2^2=\sigma$, we further get $x_2=(\sigma-\mu x_1)/(\mu-x_1)$. Now, putting this in turn into $p_1=(x_2-\mu)/(x_2-x_1)$, we obtain $p_1=(\sigma-\mu^2)/(\sigma-2\mu x_1+x_1^2)$ and hence $p_2=1-p_1=(\mu-x_1)^2/(\sigma-2\mu x_1+x_1^2)$. Therefore, $Z^*$ is given by
\begin{equation*}
\max_{x_1\in[0,\mu)}\nu(p_1H(x_1)+p_2H(x_2))=\max_{x_1\in[0,\mu)}\nu\left(\frac{\sigma-\mu^2}{\sigma-2\mu x_1+x_1^2}H(x_1)+\frac{(\mu-x_1)^2}{\sigma-2\mu x_1+x_1^2}H\left(\frac{\sigma-\mu x_1}{\mu-x_1}\right)\right)
\end{equation*}
which is exactly $\max_{x_1\in[0,\mu)}W(x_1)$.

\underline{Proof of \ref{case2}: }Let $\mathbb P^{(k)}\sim(x_1^{(k)},x_2^{(k)},p_1^{(k)},p_2^{(k)})$ be a feasible sequence with $Z(\mathbb P^{(k)})\to Z^*$. Without loss of generality let $x_1^{(k)}\leq x_2^{(k)}$. Since $p_1^{(k)}x_1^{(k)}+p_2^{(k)}x_2^{(k)}=\mu$, we must have $x_1^{(k)}\leq\mu$. Then we must have a subsequence $x_2^{(k_i)}\to\infty$, since otherwise $(x_1^{(k)},x_2^{(k)},p_1^{(k)},p_2^{(k)})$ would lie in a compact set and there would exist a subsequence $(x_1^{(k_i')},x_2^{(k_i')},p_1^{(k_i')},p_2^{(k_i')})\to(x_1^*,x_2^*,p_1^*,p_2^*)$, where $Z(\mathbb P^{(k_i')})=\nu\sum_{j=1}^2p_j^{(k_i')}H(x_j^{(k_i')})\to\nu\sum_{j=1}^2p_j^*H(x_j^*)$ by the continuity of $H$, violating the non-existence of optimal solution. By $p_1^{(k_i)}{x_1^{(k_i)}}^2+p_2^{(k_i)}{x_2^{(k_i)}}^2=\sigma$, we have $p_2^{(k_i)}=(\sigma-p_1^{(k_i)}{x_1^{(k_i)}}^2)/{x_2^{(k_i)}}^2\to0$, and $p_2^{(k_i)}x_2^{(k_i)}=(\sigma-p_1^{(k_i)}{x_1^{(k_i)}}^2)/x_2^{(k_i)}\to0$. Thus $p_1^{(k_i)}=1-p_2^{(k_i)}\to1$ and $x_1^{(k_i)}=(\mu-p_2^{(k_i)}x_2^{(k_i)})/p_1^{(k_i)}\to\mu$. Therefore,
\begin{align*}
Z(\mathbb P^{(k_i)})&=\nu\left(p_1^{(k_i)}H(x_1^{(k_i)})+p_2^{(k_i)}H(x_2^{(k_i)})\right)=\nu\left(p_1^{(k_i)}H(x_1^{(k_i)})+\frac{\sigma-p_1^{(k_i)}{x_1^{(k_i)}}^2}{{x_2^{(k_i)}}^2}H(x_2^{(k_i)})\right)\\
&\to\nu(H(\mu)+\lambda(\sigma-\mu^2))
\end{align*}

\underline{Proof of \ref{case overall}: }
First, we show that $W(x_1)\to\nu(H(\mu)+\lambda(\sigma-\mu^2))$ as $x_1\nearrow\mu$. Consider the second term of $W(x_1)$ given by
$$\lim_{x_1\nearrow\mu}\frac{\nu(\mu-x_1)^2}{\sigma-2\mu x_1+x_2^2}H\left(\frac{\sigma-\mu x_1}{\mu-x_1}\right)=\lim_{x_1\nearrow\mu}\frac{\nu(\sigma-\mu x_1)^2}{\sigma-2\mu x_1+x_2^2}\left(\frac{\mu-x_1}{\sigma-\mu x_1}\right)^2H\left(\frac{\sigma-\mu x_1}{\mu-x_1}\right)=\nu\lambda(\sigma-\mu^2)$$
and the claim follows.
Combining Parts \ref{case1} and \ref{case2} of this proposition, we must have $Z^*=\max_{x_1\in[0,\mu]}W(x_1)$.\hfill\Halmos
\endproof

\section{Proofs for Section \ref{sec:numerics data}}\label{sec:data driven proofs}
We first show a result in parallel to Theorem \ref{main reduction} for the case of \eqref{opt relaxed}:
\begin{theorem}
Suppose $h$ is bounded. Then the optimal value of \eqref{opt relaxed} is the same as
\begin{equation}
\begin{array}{ll}
\underset{\mathbb P}{\max}&\overline\nu\mathbb E[ H(X)]\\
\text{subject to\ \ }&\underline\mu\leq\mathbb E[X]\leq\overline\mu\\
&\underline\sigma\leq\mathbb E[X^2]\leq\overline\sigma\\
&\mathbb P\in\mathcal P^+
\end{array}\label{prob opt relaxed}
\end{equation}
Here the decision variable is a probability distribution $\mathbb P\in\mathcal P^+$, and $\mathbb E[\cdot]$ is the corresponding expectation. Moreover, there is a one-to-one correspondence between the feasible solutions to \eqref{opt relaxed} and \eqref{prob opt relaxed}, given by $f_+'(x+a)=\overline\nu(p(x)-1)$ for $x\in\mathbb R^+$, where $f_+'$ is the right derivative of a feasible solution $f$ of \eqref{opt relaxed} such that $f(x)=\int_a^xf_+'(t)dt+\eta$ for $x\geq a$, and $p$ is a probability distribution function that is associated with a feasible probability measure over $\mathbb R^+$ in \eqref{prob opt relaxed}.\label{thm reduction data}
\end{theorem}

\proof{Proof of Theorem \ref{thm reduction data}.}
Note that formulation \eqref{opt relaxed} can be written as
\begin{equation}
\begin{array}{lll}
\max_{\underline\beta\leq\beta\leq\overline\beta,\underline\eta\leq\eta\leq\overline\eta}&\underset{f}{\max}&\int_a^\infty h(x)f(x)dx\\
&\text{subject to\ \ }&\int_a^\infty f(x)dx=\beta\\
&&f(a)=f(a+)=\eta\\
&&f_+'(a)\geq-\overline\nu\\
&&f(x)\text{\ convex for\ }x\geq a\\
&&f(x)\geq0\text{\ for\ }x\geq a
\end{array}\label{opt relaxed1}
\end{equation}
The inner maximization is exactly \eqref{opt}, and thus by Theorem \ref{thm:prob opt} we can reformulate \eqref{opt relaxed1} as
\begin{equation*}
\begin{array}{lll}
\max_{\underline\beta\leq\beta\leq\overline\beta,\underline\eta\leq\eta\leq\overline\eta}&\underset{\mathbb P}{\max}&\overline\nu\mathbb E[H(X)]\\
&\text{subject to\ \ }&\mathbb E[X]=\frac{\eta}{\overline\nu}\\
&&\mathbb E[X^2]=\frac{2\beta}{\overline\nu}\\
&&\mathbb P\in\mathcal P^+
\end{array}
\end{equation*}
which is equivalent to \eqref{prob opt relaxed}.\hfill\Halmos
\endproof

For convenience, we denote $\widetilde{OPT}(\mathcal D)$ as the program
\begin{equation*}
\begin{array}{ll}
\underset{\mathbb P}{\max}&\overline\nu\mathbb E[ H(X)]\\
\text{subject to\ \ }&\underline\mu\leq\mathbb E[X]\leq\overline\mu\\
&\underline\sigma\leq\mathbb E[X^2]\leq\overline\sigma\\
&\mathbb P\in\mathcal D
\end{array}
\end{equation*}
where $\mathcal D$ is a collection of probability measures on $\mathbb R$. For example, \eqref{prob opt relaxed} can be written as $\widetilde{OPT}(\mathcal P^+)$. Let $\tilde Z(\mathbb P)=\overline\nu\mathbb E[H(X)]$ be the objective function in $\mathbb P$.


\begin{proposition}
The optimal value of $\widetilde{OPT}(\mathcal P^+)$ is identical to that of $\widetilde{OPT}(\mathcal P_3^+)$.\label{ext1}
\end{proposition}

\proof{Proof of Proposition \ref{ext1}.}
For $\mathbb P$ feasible in $\widetilde{OPT}(\mathcal P^+)$, let $\mu=\mathbb E[X]$ and $\sigma=\mathbb E[X^2]$ be its first and second moments. By Theorem \ref{equivalence finite} there must exist $\mathbb P'\in\mathcal P_3^+$ with the corresponding expectations $\mathbb E'[X]=\mu$ and $\mathbb E'[X^2]=\sigma$ such that $\tilde Z(\mathbb P)\leq\tilde Z(\mathbb P')$. Since $\mathcal P_3^+\subset\mathcal P^+$, we conclude that the optimal value of $\widetilde{OPT}(\mathcal P^+)$ is identical to that of $\widetilde{OPT}(\mathcal P_3^+)$.\hfill\Halmos
\endproof

\proof{Proof of Theorem \ref{main thm relaxed}.}
Theorem \ref{main thm relaxed} follows from Theorem \ref{thm reduction data} and Proposition \ref{ext1}, in the same way as the proof of Theorem \ref{main thm}.\hfill\Halmos
\endproof

\begin{proposition}
Under Assumption \ref{regularity H}, $\widetilde{OPT}(\mathcal P^+)$ has the same optimal value as $\widetilde{OPT}(\mathcal P_2^+)$.\label{simplified1}
\end{proposition}

\proof{Proof of Proposition \ref{simplified1}.}
We know from Proposition \ref{ext1} that $\widetilde{OPT}(\mathcal P^+)$ has the same optimal value as $\widetilde{OPT}(\mathcal P_3^+)$. Any $\mathbb P$ feasible for $\widetilde{OPT}(\mathcal P_3^+)$ must necessarily have bounded support, say on $[0,M]$. By Lemma \ref{2-supp prelim} there must exist $\mathbb P'\in\mathcal P_2^+$, with the same first and second moments as $\mathbb P$, such that $\tilde Z(\mathbb P)\leq\tilde Z(\mathbb P')$. Since $\mathcal P_2^+\subset\mathcal P_3^+$, this implies that $\widetilde{OPT}(\mathcal P_3^+)$ has the same optimal value as $\widetilde{OPT}(\mathcal P_2^+)$, which concludes the proposition.\hfill\Halmos
\endproof

The following explains the origin of the two subproblems in \eqref{new opt}:

\begin{lemma}
Under Assumption \ref{regularity}, and let $\overline\sigma\geq\underline\mu^2$. The optimal value of $\widetilde{OPT}(\mathcal P_2^+)$ is given by $\tilde Z^*=\max\{\tilde Z_1^*,\tilde Z_2^*\}$, where $\tilde Z_1^*$ is the optimal value of
\begin{equation}
\begin{array}{ll}
\underset{\mathbb P}{\max}&\overline\nu\mathbb E[H(X)]\\
\text{subject to\ \ }&\mathbb E[X]=\overline\mu\\
&\underline\sigma\leq\mathbb E[X^2]\leq\overline\sigma\\
&\mathbb P\in\mathcal P_2^+
\end{array}\label{subopt1}
\end{equation}
and $\tilde Z_2^*$ is the optimal value of
\begin{equation}
\begin{array}{ll}
\underset{\mathbb P}{\max}&\overline\nu\mathbb E[H(X)]\\
\text{subject to\ \ }&\underline\mu\leq\mathbb E[X]\leq\overline\mu\\
&\mathbb E[X^2]=\overline\sigma\\
&\mathbb P\in\mathcal P_2^+
\end{array}\label{subopt2}
\end{equation}
respectively. 
\label{subprograms}
\end{lemma}

\proof{Proof of Lemma \ref{subprograms}.}
We argue that to solve $\widetilde{OPT}(\mathcal P_2^+)$, it suffices to restrict attention to the feasible region $\{\mathbb P\in\mathcal P_2^+:\mathbb E[X]=\overline\mu,\underline\sigma\leq\mathbb E[X^2]\leq\overline\sigma\}\cup\{\mathbb P\in\mathcal P_2^+:\underline\mu\leq\mathbb E[X]\leq\overline\mu,\mathbb E[X^2]=\overline\sigma\}$. Since $h\geq0$, $\tilde Z^*\geq0$. There is nothing to prove if $\tilde Z^*=0$. So suppose $\tilde Z^*>0$. There exists $\mathbb P\sim(x_1,x_2,p_1,p_2)\in\mathcal P_2^+$ with one of the $x_i$'s having $ H(x_i)>0$ and $p_i>0$. Now suppose $\mathbb P$ satisfies $\mathbb E[X]<\overline\mu$ and $\mathbb E[X^2]<\overline\sigma$. We can increase $x_i$ so that $\mathbb E[X]\leq\overline\mu$ and $\mathbb E[X^2]\leq\overline\sigma$ remain satisfied, and $\tilde Z^*(\mathbb P)$ is at least as large as before since $ H(x)$ is non-decreasing. Hence any $\mathbb P$ such that $\mathbb E[X]<\overline\mu$ and $\mathbb E[X^2]<\overline\sigma$ must have $\tilde Z(\mathbb P)\leq\tilde Z(\mathbb P')$ for some $\mathbb P'\in\{\mathbb P\in\mathcal P_2^+:\mathbb E[X]=\overline\mu,\underline\sigma\leq\mathbb E[X^2]\leq\overline\sigma\}\cup\{\mathbb P\in\mathcal P_2^+:\underline\mu\leq\mathbb E[X]\leq\overline\mu,\mathbb E[X^2]=\overline\sigma\}$. 
This proves the lemma.\hfill\Halmos
\endproof

\proof{Proof of Theorem \ref{main thm simplified relaxed}.}
Lemma \ref{subprograms} allows one to consider only the programs \eqref{subopt1} and \eqref{subopt2} when solving $\widetilde{OPT}(\mathcal P_2^+)$. Theorem \ref{main thm simplified relaxed} then follows from Lemma \ref{consistency relaxed}, Theorem \ref{thm reduction data} and Proposition \ref{simplified1}, using the same line of arguments in the proof of Theorem \ref{main thm simplified}.\hfill\Halmos
\endproof

\end{document}